\documentclass[11pt]{article}
\usepackage{amsmath,amssymb,color,epsfig,cite}
\usepackage{multirow}

\usepackage{amsfonts}
\newcommand{\hoch}[1]{$\, ^{#1}$}

\makeatletter
\@addtoreset{equation}{section}
\makeatother

\newtheorem{theorem}{Theorem}

\newcommand{\be}{\begin{equation}}
\newcommand{\ee}{\end{equation}}
\newcommand{\bea}{\setlength\arraycolsep{2pt} \begin{eqnarray}}
\newcommand{\eea}{\end{eqnarray}}
\newcommand{\nn}{\nonumber}

\newcommand{\bpm}{\begin{pmatrix}}
\newcommand{\epm}{\end{pmatrix}}

\newcommand{\td}{\mathrm{d}}
\newcommand{\te}{\mathrm{e}}

\def\ft#1#2{{\textstyle{\frac{\scriptstyle #1}{\scriptstyle #2} } }}
\def\fft#1#2{{\frac{#1}{#2}}}

\def\0{{\sst{(0)}}}
\def\1{{\sst{(1)}}}
\def\2{{\sst{(2)}}}
\def\3{{\sst{(3)}}}
\def\4{{\sst{(4)}}}
\def\5{{\sst{(5)}}}
\def\6{{\sst{(6)}}}
\def\7{{\sst{(7)}}}
\def\8{{\sst{(8)}}}
\def\sst#1{{\scriptscriptstyle #1}}

\begin{document}


\begin{center}
{\large {\bf Superradiant Instability of Extremal Black Holes in STU Supergravity}}

\vspace{10pt}
Zhan-Feng Mai\hoch{1}, Run-Qiu Yang\hoch{1}, H. L\"{u}\hoch{1,2}

\vspace{10pt}

\hoch{1} {\it Center for Joint Quantum Studies and Department of Physics\\ School of Science, Tianjin University, Tianjin 300350, P.R.~China}

\vspace{10pt}

\hoch{2} {\it Peng Huangwu Center for Fundamental Theory, Hefei, Anhui 230026, P.R.~China}

\vspace{40pt}

\underline{ABSTRACT}

\end{center}

We investigate the superradiant (in)stability of the extremal multi-charge static black holes in the STU supergravity model, which reduce to the RN black hole when all the charges are equal.
We first show that the frequency of quasi-bound states is necessarily complex and obtain the corresponding superradiant condition. We then study the effective potential of the Schr\"odinger-like equation associated with the radial function of the charged scalar field. We find that trapping-well configurations can emerge with either a single peak or double peaks.  We numerically obtain the corresponding unstable quasi-bound states, organized under the overtone number, as well as how the charged black holes deviate from the RN black hole. We find that the STU black holes are superradiantly unstable as long as not all the charges are equal, indicating that the superradiant stability of the extremal RN black hole is a fine-tuning result in the framework of the STU supergravity model.

\vfill {\footnotesize  zhanfeng.mai@gmail.com\ \ \ aqiu@tju.edu.cn\ \ \ mrhonglu@gmail.com}

\thispagestyle{empty}

\pagebreak

\tableofcontents
\addtocontents{toc}{\protect\setcounter{tocdepth}{2}}


\newpage

\section{Introduction}

With the detection of gravitational waves from two merging black holes and photo images of black holes \cite{LIGOScientific:2016aoc,EventHorizonTelescope:2019dse}, an increasing number of evidences suggest that black holes, predicted by Einstein's theory of general relativity, exist in our Universe. An important method to probe the black holes is to investigate the linear perturbations of matter or gravitational fields in the black hole background. These perturbations are typically described by second-order differential equations.
The nature of the black hole horizon requires that the boundary conditions of these linearized fields must be ingoing. We must also require that the fields are all finite at the asymptotic infinity. This leads to three inequivalent classes of asymptotic boundary conditions:
\begin{itemize}

\item Scattering processes: there are both ingoing and outgoing modes.

\item Quasi-normal modes: there are outgoing modes only.

\item Quasi-bound states: the fields vanish.

\end{itemize}
Specifically for a massive scalar particle field of mass $m_p$, if its frequency is sufficiently high ($\omega>m_p$) so that it is free and wavelike in the asymptotic infinity, one can either study the scattering process by not imposing any boundary condition, or study the quasinormal modes which reveal the (in)stability of the system by imposing only the outgoing boundary condition at infinity. When $\omega<m_p$, the wave function either exponentially diverge or converge and the quasi-bound states are the latter case.

In a scattering process, we typically expect that the amplitude of an ingoing (incident) mode is bigger than that of the corresponding outgoing (reflection) mode. However, the opposite superradiance effect can arise in rotating or charged black holes \cite{Press:1972zz,Bekenstein:1973mi,Teukolsky:1974yv,Starobinskil:1974nkd,Starobinsky:1973aij}, indicating that the scattering process can actually extract charges and/or angular momenta, and hence the energies from  black holes. This phenomenon occurs for low-frequency excitations with $\omega<\omega_c$ where $\omega_c$ is a certain critical value that depends on the mass, angular momentum and electric charge of the black hole. (See a review on superradiance\cite{Brito:2015oca}.)

A brutal force implementation of the third boundary condition was first introduced by Press and Teukolsky \cite{Press:1972zz}. They considered putting a mirror outside the black hole horizon such that the fields would all vanish asymptotically.  It turns out the bounded scalar perturbation of a Kerr black hole can be exponentially amplified by extracting energy from the black hole, leading to a bomb effect. This black hole bomb mechanism was later realized naturally using massive scalar fields since they fall off exponentially at the asymptotic infinity for low-frequency excitations, leading to the superradiant instability of the Kerr or Kerr-Newmann black hole \cite{Damour:1976kh,Detweiler:1980uk,Furuhashi:2004jk,Dolan:2007mj,Huang:2018qdl}. Interestingly, for asymptotic AdS spcaetimes, the AdS boundary can provide such a ``mirror'' and superradiant instability can indeed arise in both charged static and rotating black holes\cite{Cardoso:2004hs,Cardoso:2004nk,Li:2012rx,Wang:2014eha,Bosch:2016vcp,Ganchev:2016zag}. In this paper, we shall consider only four-dimensional black holes that are asymptotic to the Minkowski spacetime.

Typically a black hole has three conserved quantities, the mass $M$, angular momentum $J$ and electric charge $Q$. For scalar field perturbations, it can be easily established that there is no superradiant effect or instability of neutral static black holes such as the Schwarzschild black hole.  The superradiant effects were shown to arise when $J Q\ne 0$. There is hitherto no example of superradiant instability when $J=0$.  Recently, it was shown that the Reissner-Nordstr\"{o}m (RN) black hole is superradiantly stable \cite{Hod:2012wmy,Huang:2015jza,Hod:2015hza}. Particularly, in \cite{Hod:2012wmy}, Hod examined the two necessary conditions for the existence of superradiant instability: a) there exists an $\omega_c>0$ such that we can impose the superradiant condition $\omega<\omega_c$. b) there exists a potential well to trap the quasi-bound states.  It was shown that the two conditions could not be simultaneously satisfied for the extremal RN black hole.

In this paper, we study the superradiant instability of the RN black hole from the perspective of more fundamental theories such as strings and supergravities, where the extremal RN black hole emerges as a special BPS state. Specifically we consider STU supergravity, which is a special class of Einstein-Maxwell-dilaton theories involving four Maxwell fields and three dilaton scalars \cite{Duff:1995sm}. The theory admits a general class of four-charge black hole solutions and they reduce to the RN black hole when all the charges are exactly equal.  The BPS (no-force) condition implies that in the extremal limit, the RN black hole can be viewed as a bound state with threshold binding energy of the four more fundamental ingredients that have origins as strings and branes in $D=10$ or 11 dimensions, e.g. \cite{Cvetic:1999xp}. We consider an additional massive scalar field charged under all the four Maxwell fields and we find that superradiant condition $\omega<\omega_c$ can be satisfied for the generic black hole charge configuration. When all the four black hole charges are equal, it can be demonstrated that there is no potential well when $\omega<\omega_c$. However, we find that a potential well can always form as long as the four charges of the extremal black hole are not all exactly the same. Furthermore, our numerical analysis indicates that the extremal STU black holes can all be superradiantly unstable provided that the black hole charges are not the same, regardless how small the difference is. Our results are the first in literature where the superradiant instability is demonstrated in spherically-symmetric and static black holes.

The paper is organized as follows. In Sec.~\ref{modset}, we briefly introduce the STU supergravity model and the associated extremal charged black hole with four $U(1)$ charges. By introducing a charged scalar perturbation, we obtain the radial wave equation, giving the boundary condition for the quasi-bound states. In Sec.~\ref{cond}, we illustrate three necessary conditions for superradiant instability. We then analyse the conserved flux near the horizon for the superradiant condition and point out the equivalence between superradiant instability and unstable quasi-bound states. We give a no-go theorem of superradiant instability for the extremal RN black hole that emerges from four equal charges. In Sec.~\ref{numer}, we numerically construct the quasi-bound states by the shooting method and analyse the superradiant instability spectrum. Our results indicate that the extremal STU black holes
are unstable as long as not all the charges are equal. In Sec.~\ref{conclu}, the conclusion and some further discussions are given. In Appendix A, we analyse the asymptotic boundary conditions of the radial wave function in tortoise coordinates.  In appendix B, we prove that the frequencies of quasi-bound states and also quasinormal modes are necessarily complex.  We put most of plots and graphs of numerical results in appendix C.

\section{Model Setup}\label{modset}

The STU supergravity model in four dimensions is the low-energy effective theory of six-dimensional string reduced on 2-torus \cite{Duff:1995sm}. The minimal field content of the bosonic sector consists of the metric $g_{\mu\nu}$ and four Maxwell fields $(\tilde A_1, \tilde A_2, \tilde A_3, \tilde A_4)$, in additional to the three dilatonic scalars and three axions that form three $SL(2,\mathbb R)$ complex scalars. For our purpose of studying the properties of electrically-charged and static black holes, the axions decouple and the relevant Lagrangian is (e.g. \cite{Cvetic:1999xp})
\begin{eqnarray}
{\cal L}&=& \sqrt{-g} \Big(R -\fft12(\partial\vec\varphi)^2 - \fft14 \sum_{i=1}^4 e^{\vec a_i \cdot \vec \varphi} \tilde F_i^2\Big)\,,\nn\\
\vec a_1 &=& = (1, 1, 1),\qquad\vec a_2 = (1, -1, -1),\qquad \vec a_3 = (-1, 1, -1)\,,\nn\\
\vec  a_4 &=& (-1, -1, 1),\qquad \vec \varphi=(\varphi_1,\varphi_2,\varphi_3)\,,\qquad \tilde F_i=d\tilde A_i\,.
\end{eqnarray}
The theory admits an exact solution of static charged black holes. In the extremal limit, the solution is
\begin{eqnarray}\label{4chargesol}
ds^2 &=& - (\tilde H_1\tilde H_2 \tilde H_3 \tilde H_4)^{-\fft12} dt^2 + (\tilde H_1\tilde H_2 \tilde H_3 \tilde H_4)^{\fft12}(dr^2 + r^2 d\Omega_2^2)\,,\nn\\
e^{\ft12\vec a_i \cdot \vec \varphi} &=& \tilde H^{-1}_i(\tilde H_1\tilde H_2\tilde H_3\tilde H_4)^{\fft14}\,,\qquad \tilde A_i=(\tilde H_i^{-1} -1) dt\,,\qquad \tilde H_i = 1 + \fft{4\tilde Q_i}{r^2}\,.
\end{eqnarray}
where $\tilde Q_i$ are the electric charges.  The mass and entropy of these extremal black hole
are
\be
M=\tilde Q_1 + \tilde Q_2 + \tilde Q_3 +\tilde Q_4\,,\qquad S=16\pi \sqrt{\tilde Q_1 \tilde Q_2 \tilde Q_3 \tilde Q_4}\,.
\ee
Note that $\tilde Q_i$'s here are nonnegative. When the electric charges $\tilde Q_i$ are all equal, the scalars all decouple and the solution becomes the extremal RN black hole.  Thus in string theory, the RN black hole is not a fundamental object, but a bound state with threshold binding energy of fundamental ingredients that have higher dimensional origins as strings and branes.

The STU model can be further truncated to involve only one scalar and two Maxwell fields.  Two situations can arise.  One is to set $\tilde A_1=\tilde A_2$ and $\tilde A_3=\tilde A_4$ and the other is to set $\tilde A_2=\tilde A_3=\tilde A_4$.  In both cases, the relevant Lagrangian can be cast into \cite{Lu:2013eoa}
\begin{equation}\label{action}
{\cal L}= \sqrt{-g} \left(R - \frac{1}{2}(\partial \phi)^2-\frac{1}{4}\te^{\alpha_1 \phi}F_1^2-\frac{1}{4}\te^{\alpha_2 \phi}F_2^2  \right) \, ,
\end{equation}
where $F_1= \td A_1$ and $F_2=\td A_2$. The dilaton coupling constants $(\alpha_1, \alpha_2)$ satisfy
\begin{equation}\label{N1N2}
\alpha_1 \alpha_2 = -1 \, , \quad N_1 \alpha_1 + N_2 \alpha_2 = 0 \, ,  \quad N_1 + N_2 = 4
\end{equation}
The $(N_1,N_2)=(2,2)$ and $(1,3)$ cases reduce to the two above special simplified STU models respectively. Charged black holes for generic $(N_1,N_2)$ were obtained in \cite{Lu:2013eoa} and the extremal solutions are
\begin{eqnarray}\label{exbg}
&&\td s^2=-\left(H_1^{N_1}H_2^{N_2} \right)^{-\frac{1}{2}}\td t^2 +(H_1^{N_1} H_2^{N_2})^{\frac{1}{2}}\left(\td r^2 + r^2 \td \Omega^2_{2} \right)\,, \cr
&& A_i = N_i \left(H_i^{-1} -1  \right) \td t \,,\qquad \phi =\sum_{i=1}^2 \frac{1}{2}\alpha_i N_i \log H_i\, ,\qquad H_i =1+ \fft{4Q_i}{\sqrt{N_i}\,r}\,.
\end{eqnarray}
Note that when $ \frac{Q_1}{\sqrt{N_1}} = \frac{Q_2}{\sqrt{N_2}} $, the above reduces to the extremal RN black hole, as we have discussed earlier. The truncated two-charge system is much simpler than the full STU model, while keeping the essential feature that the RN black hole is a bound state.

In order to study the superradiant instability, we consider a larger theory with additional massive scalar $\Phi$, charged under all the four $U(1)$ fields. The linear perturbation of this scalar is governed by its charged Klein-Gordon equation
\begin{equation}\label{kg1}
\left(g^{\mu\nu} D_\mu D_\nu - m_p^2\right) \Phi=0 \, , \quad \quad D_{\mu} := \nabla_\mu - i \tilde q_1 \tilde A_{1 \mu} - i \tilde q_2\tilde A_{2 \mu} - i\tilde q_3\tilde A_{3 \mu}- i\tilde q_4 \tilde A_{4 \mu}\,,
\end{equation}
in the background \eqref{4chargesol}, where $(m_p,\tilde q_i)$ denote the fundamental mass and charges of the scalar. While we have $m_p>0$, the fundamental charges $\tilde q_i$'s can be positive or negative, associated with particles or antiparticles. By contrast, $\tilde Q_i$'s are all positive. Since the background is spherically symmetric, static and electrically charged,  the general solution can be expressed as linear superpositions of different frequency modes:
\begin{equation}\label{anz}
\Phi = \te^{- i \omega t} R(r) {\rm Y}_{\ell m}\left(\theta, \varphi\right)\,,
\end{equation}
where $Y_{\ell,m}$'s are spherical harmonics. The KG equation now reduces to
\begin{equation}\label{radeq}
-r^2 \frac{\td }{ \td r}\left( r^2 \frac{\td R}{\td r}  \right) + U R = 0 \, ,
\end{equation}
with
\begin{eqnarray}
 U(r)=  \ell(\ell+1)r^2+ m_p^2 r^4 \sqrt{\tilde H_1\tilde H_2 \tilde H_3 \tilde H_4}
 -r^4 \left(\omega + \sum_{i=1}^4 \tilde q_i \left(\tilde H_i^{-1} -1  \right) \right)^2  \tilde H_1\tilde H_2 \tilde H_3 \tilde H_4\,.
\end{eqnarray}
Note that only $\ell$, not $m$, of $Y_{\ell,m}$ enters the reduced wave equation. In the regions of both the near horizon ($r \to 0$) and infinity ($r \to \infty$), the radial wave function $R(r)$ can be solved asymptotically:
\begin{equation}\label{asy1}
R(r)|_{r \to 0} \sim \te^{i\frac{\#(\omega - \omega_c)}{r}} \, , \quad  R(r)|_{r \to \infty} \sim \frac{\te^{-\sqrt{m_p^2 - \omega^2 }\, r} }{r}\, ,
\end{equation}
where
\begin{equation}
\omega_c \equiv \tilde q_1 + \tilde  q_2 + \tilde q_3 + \tilde q_4 \,.\label{omegac}
\end{equation}
Note that we have chosen the ingoing boundary condition on the horizon so that the coefficient ``$\#$'' of $\omega$ is positive.  The quantity $\omega_c$ is certain critical value that determines whether the black hole energy can be extracted; we shall discuss this presently. Here $\omega_c$ depends only on the fundamental charges; it is the sum of them all. At the asymptotic infinity, when $m_p<\omega$, the solution is wavelike, describing scattering processes in general, or quasinormal modes if we impose further the outgoing boundary condition asymptotically.

In this paper, we focus on the quasi-bound states where
\begin{equation}\label{cond1}
\omega<m_p\,,
\end{equation}
such that the wave function falls off exponentially at the asymptotic infinity. It has a consequence, as in the case of quasinormal modes, that the frequency must be discrete.  Furthermore, as we show in appendix \ref{qbmcom}, the frequency must be complex, namely
\begin{equation}
\omega_{\rm \sst{QBS}}=\omega_r + {\rm i} \omega_i\,,
\end{equation}
where $\omega_r$ and $\omega_i$ denote the real and  imaginary parts of $\omega_{\rm QBS}$ respectively. The existence of such bound states with positive $\omega_i$ indicates that they will grow exponentially with time and therefore the system is unstable at the linear level. The focus of this paper is to analyse this superradiant instability for the charged black holes in the STU model.

\section{Superradiant condition for complex frequencies}\label{cond}

There are important differences between quasi-bound states(or quasinormal modes) and those associated with the scattering processes.  In the latter case, the frequency $\omega$ is real, whilst the former typically involves complex frequencies.  It is thus important to obtain the corresponding superradiant conditions of the two different situations.

\subsection{Superradiant condition for scattering processes}

We first review the superradiant condition for the scattering processes. As a concrete example, we illustrate this by using the extremal black hole of the STU model. The superradiant scattering is typically discussed in the tortoise coordinates, which we give in the appendix \ref{tort}.  For the appropriately scaled radial wave function \eqref{barR}, we have, with $m_p<\omega$,
\begin{equation}\label{asyso1}
\bar {R} = \left\{
\begin{aligned}
& {\cal T} \te^{-i (\omega - \omega_c)y}  \quad \quad  &&y \to -\infty \\
& {\cal I} \te^{-i \sqrt{\omega^2-m_p^2}y} + {\cal R} \te^{i \sqrt{\omega^2-m_p^2}y} \quad \quad && y \to \infty \, ,
\end{aligned}
\right.
\end{equation}
where $\omega_c$ for the extremal STU black holes is given by \eqref{omegac}, and ${\cal I}, {\cal R}$ and ${\cal T}$ denote the incident, reflection and transmission coefficients respectively. The conservation of the Wroskian determinant associated with the radial wave function
\begin{equation}
W|_{y \to - \infty} = W|_{y \to \infty} \, , \quad \text{where} \quad W=\bar R^{*}\frac{\td \bar R}{\td y} - \bar R \frac{\td \bar R^{*}}{\td y}\, ,
\end{equation}
implies that
\begin{equation}\label{Wros}
|{\cal R}|^2 = |{\cal I}|^2-\frac{\omega - \omega_c}{\sqrt{\omega^2-m_p^2}}|{\cal T}|^2  .
\end{equation}
Thus when $\omega < \omega_c$, the reflecting amplitude is larger than the incident amplitude, namely $|{\cal R}|^2 > |{\cal I}|^2 $. It is important to note that in this scattering process, the frequency $\omega$ is real. The situation changes for quasi-bound states where $\omega$ is complex and we shall discuss this in the next subsection.

\subsection{Superradiant condition for quasi-bound states}

As shown in appendix \ref{qbmcom}, the frequency of the quasi-bound states is necessarily complex. The superradiant condition of the previous subsection is thus nonapplicable. In order to derive the new condition, we followed the observation \cite{DiMenza:2014vpa} that the energy momentum tensor associated with the charged scalar perturbation do not conserved, but there still exists an energy conserved current.  Specially, the energy momentum tensor associated with the complex scalar perturbation gives
\begin{equation}\label{T1}
T^{\Phi}_{\mu\nu}=D_{(\mu}\Phi D^{\dag}_{\nu )} \Phi^{\dag}  - \frac{1}{2}g_{\mu\nu}\left(D^\rho \Phi D^{\dag}_\rho \Phi^{\dag}- m_p^2 \Phi \Phi^{\dag}\right) \,.
\end{equation}
However, it is straightforward to verify that
\begin{equation}
\nabla^\mu T^{\Phi}_{\mu\nu}= \left(\sum^{4}_{k=1}  q_k \tilde F_{k \nu\rho}\right)\left(\sum^4_{j=1} \Phi^{\dag}\Phi\,\tilde q_j \tilde A^\rho_j +i \left(\Phi^\dag \nabla^\rho \Phi - \Phi \nabla^\rho \Phi^{\dag}\right) \right) \ne 0 \, .
\end{equation}
The reason why it is not conserved is that at the perturbation level, the source term contributed by four $U(1)$ fields $(\tilde A_1, \tilde A_2, \tilde A_3, \tilde A_4)$ has been neglected. Following the method given in \cite{DiMenza:2014vpa}, we find a new ``energy momentum tensor''
\begin{equation}
\tilde{T}_{\mu\nu} = \sum_{j=r,i} 2\nabla_\mu \Phi_j \nabla_\nu \Phi_j  -  g_{\mu\nu}\left(\nabla^\rho \Phi_j \nabla_\rho \Phi_j +  \left(\sum^4_i \tilde q_i \tilde A^\rho_i \right)^2  \Phi_j^2 \right) - m_p^2\Phi^2_j \, ,
\end{equation}
where $\Phi_r = \frac{1}{2} \left( \Phi + \Phi^{\dag} \right)$ and $\Phi_i = \frac{1}{2 i} (\Phi - \Phi^{\dag})$. Although this new energy momentum tensor does not satisfy the conservation law either, namely $\nabla^\mu \tilde{T}_{\mu\nu} \not = 0 $, there does exist a conserved energy current that satisfies the conservation law, namely
\begin{equation}\label{cons}
J^\mu_{\rm E} = \tilde{T}^\mu{}_\nu \left(\frac{\partial}{\partial t} \right)^\nu\,,\qquad
\nabla_\mu J^\mu _{\rm E} = 0 \,.
\end{equation}
In the case of the quasi-bound states with complex frequencies, we can read off the growth rate of total energy outside of the horizon from $J^{\mu}_{\rm E}$ that
\begin{equation}\label{dE1}
\frac{\partial  E}{\partial  t} = 2 \omega_ i E  \, , \quad \text{where} \quad E = \int_{V} \sqrt{-g}J^0_{\rm E} \, .
\end{equation}
The growth rate of the total energy outside the horizon depends on the sign of the imaginary part of the quasi-bound state. A state is stable when its $\omega_i\le 0$, but unstable when $\omega_i>0$.

We now derive a necessary condition for the superradiant instability. The conservation law Eq.~\eqref{cons} implies that
\begin{equation}\label{Gauss}
\left. \frac{\partial }{\partial t} \int_V J^{0}_{\rm E}  = C_{l m}\, J^r_{\rm E} \right|^{r \to \infty}_{r \to 0} \, ,
\end{equation}
where $C_{lm} = \int Y_{lm} Y^{*}_{l m} \sin \theta \td \theta \td \varphi = \frac{4 \pi}{2l+1}$. Recall that the radial wave function decays exponentially and it gives a vanishing energy current at the spatial infinity. Eq.~\eqref{dE1} and Eq.~\eqref{Gauss} give
\begin{equation}
2E \omega_i \varpropto - (|\omega|^2 - \omega_c\omega_r) \, ,
\end{equation}
where the proportionality coefficient is some positive numerical number. We therefore conclude that a necessary condition is
\begin{equation}
\frac{|\omega|^2}{\omega_r} < \omega_c \, .\label{earliercond}
\end{equation}
This condition however is not restrictive enough to guide our numerical calculation since it involves also the unknown $\omega_i$.  When $\omega_i=0$, the above condition reduces to the one of scattering process discussed earlier. However, for complex $\omega$, we can do better.
This is because the superradiant condition derived from the energy current does not give any detail constraints on the conserved charge. The linear perturbation has also a charged conserved current defined as the Klein-Gordon product
\begin{equation}
i I_Q^\mu = \Phi^\dag D^\mu \Phi - \Phi (D^\mu \Phi)^{\dag} \, ,
\end{equation}
which satisfies the charged conservation law, $\nabla_\mu I^\mu = 0 $. Similar to the case of energy flux, the growth rate of total charge outside the horizon can be given as
\begin{equation}
\frac{\partial Q}{\partial t} = 2 \omega_i Q \, , \quad \text{where} \quad Q = \int_{V} \sqrt{-g} I_Q^0 \, .
\end{equation}
Note that this is a global charge, not the black hole electric charges $\tilde Q_i$ or the scalars' fundamental charges $q_i$'s. The growth rate of the total charge depends on the sign of $\omega_i$, whose positive values imply instability.

To obtain the condition for the associated superadiant condition, we adopt the charge conservation law $\nabla_\mu I^\mu = 0$ and the Gauss law, giving
\begin{equation}
\left. \frac{\partial Q}{\partial t} = C_{l m } \int_V  I_Q^r \right|^{r \to \infty}_{r \to 0} \, .
\end{equation}
Due to the exponentially decay of the charged scalar leading to vanishing charged current at the spatial infinity, we arrive at
\begin{equation}
2Q\omega_i  \varpropto -(\omega_r -\omega_c )\,.
\end{equation}
For unstable states, we must have positive $\omega_i$, we thus obtain a practical condition for superradiant instability:
\begin{equation}\label{cond2}
\omega_r < \omega_c \,.
\end{equation}
Note that this condition automatically implies the earlier condition \eqref{earliercond}.
Since the quasi-bound states also require that $\omega<m_p$, the above condition does not constrain the relation between the scalar's mass $m_p$ and its fundamental charge $q$ that determines $\omega_c$. On the other hand, for scattering processes or quasinormal modes, the condition $\omega>m_p$ implies that the scalar must satisfy $m_p<q$ for it to experience the superradiant effect.

\subsection{Trapping well condition}

\begin{figure}[htpb]
  \centering
  \includegraphics[width=0.45\textwidth]{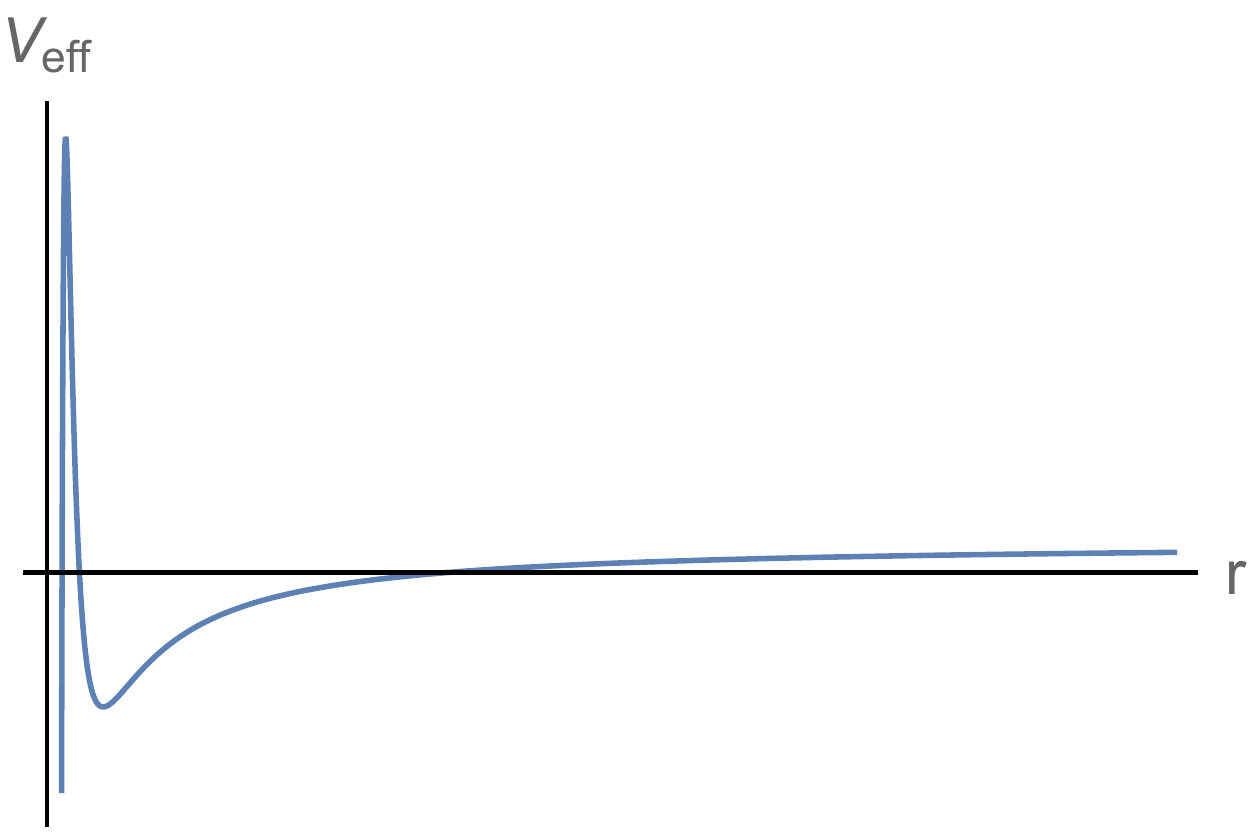}
  \includegraphics[width=0.45\textwidth]{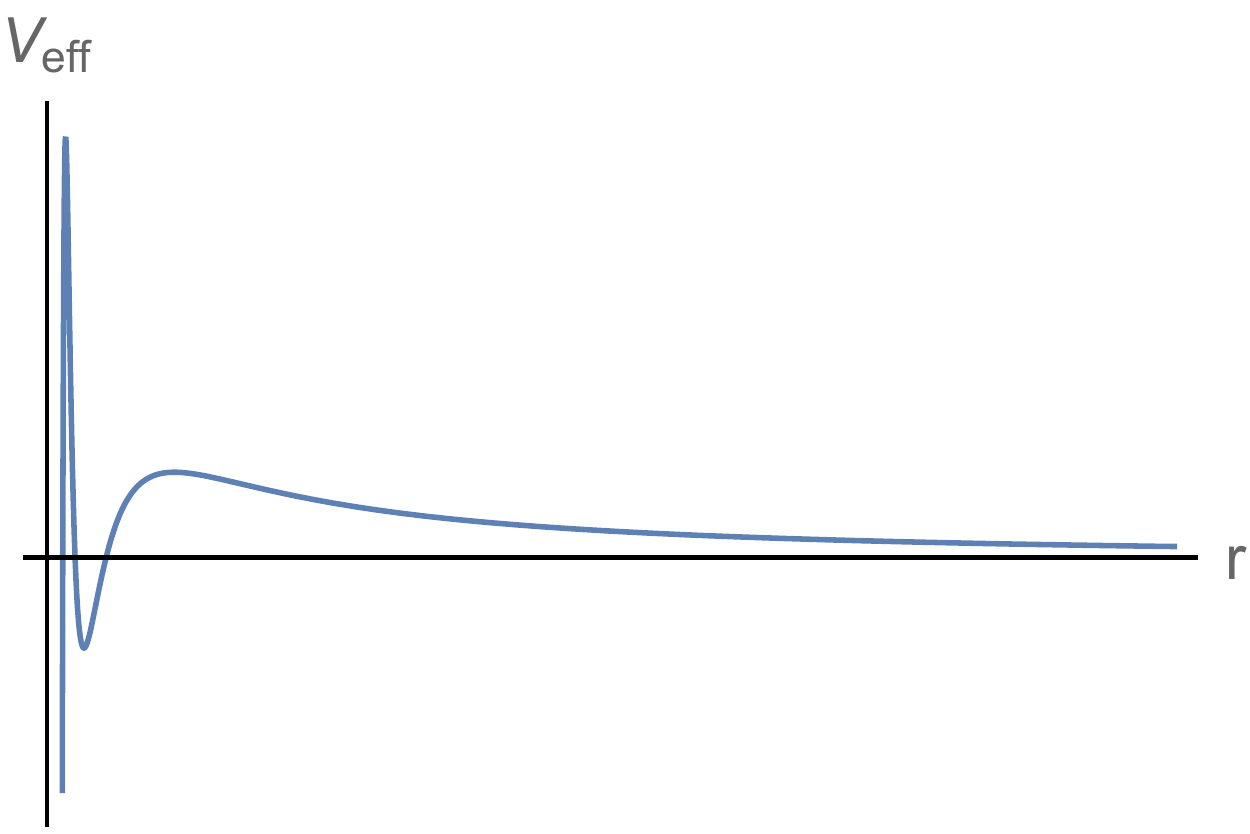}

 \caption{Two types of trapping wells can arise in $V_{\rm eff}$. The left has one peak the well lies between the peak and the asymptotic infinity. The right has two peaks with the well sandwiched between.}\label{veffshapes}
\end{figure}

In order to discuss superradiant instability, it is necessary to construct first the quasi-bound states.  To study the existence of such states, it is instructive to cast the equation \eqref{anz} in terms of the Schr\"odinger-like equation.  To do so, we define $\tilde R=r R$ such that the function $\tilde R$ satisfies the Schr\"odinger-like equation
\begin{equation}
-\fft{d^2\tilde R}{dr^2} + V_{\rm eff} \tilde R = \omega^2 \tilde R\,,\label{schrlike}
\end{equation}
where $r\in (0,\infty)$ and the potential for the general four-charge background \eqref{4chargesol} is
\begin{eqnarray}
V_{\rm eff} &=& \omega^2 + \fft{U}{r^4}\cr
&=&\omega^2 + m_p^2 \sqrt{\tilde H_1 \tilde H_2 \tilde H_3 \tilde H_4} -
\left(\omega-\sum_{i=1}^4 \fft{4\tilde q_i \tilde Q_i}{r\tilde H_i} \right)^2\tilde H_1 \tilde H_2 \tilde H_3 \tilde H_4 + \fft{\ell(\ell+1)}{r^2}\,.\label{4chargeveff}
\end{eqnarray}
Although it is not quite the same as the Schr\"odinger equation since the eigenvalue $\omega^2$ appears in the potential $V_{\rm eff}$ as well, we expect that a bound state arises only when its corresponding potential has a trapping well. To study the shape of $V_{\rm eff}(r)$, we note that
\be
V_{\rm eff} \sim
\left\{
  \begin{array}{ll}
    \fft{\alpha}{r^4}, &\qquad r\rightarrow 0\,; \\
    m_p^2 + \fft{\beta}{r}, &\qquad r\rightarrow \infty\,.
  \end{array}
\right.\label{vefffalloff}
\ee
with
\be
\alpha=-256(\omega-\sum_{i=1}^4 \tilde q_i)^2 \prod_{i=1}^4 \tilde Q_i\,,\qquad
\beta = 8 \sum_{i=1}^4 \tilde q_i \tilde Q_i + 2 (m_p^2 - 2\omega^2) \sum_{i=1}^4 \tilde Q_i\,.
\ee
Thus we see that $\alpha<0$, but $\beta$ can be both positive or negative.
The negative $\alpha$ is consistent with the fact that $r=0$ horizon is not the end of the space. It is a coordinate singularity, not an infinite potential barrier; it describes the other asymptotic region in the tortoise coordinates. For positive $\beta$, there can be at least one maximum for which $V_{\rm eff}$ is positive.  In general, there can be odd numbers of extrema and the existence of a trapping well requires at least three extrema. For negative $\beta$, there can be no extremum, or an even number of extrema. We find trapping wells with two extrema.  We illustrate the shapes of $V_{\rm eff}(r)$ of the two different types of trapping wells in Fig.~\ref{veffshapes}.

The above discussion assumed that $\omega$ was real. However, as we show in appendix \ref{qbmcom}, the frequency for a quasi-bound state is necessarily complex. Thus we have in general
\be
V_{\rm eff} = V_{\rm eff}^r + i V_{\rm eff}^i\,,\qquad
\omega^2 = (\omega^2)_r + i (\omega^2)_i\,.
\ee
The real and complex parts of the Schr\"odinger-like equation \eqref{schrlike} are
\bea
&&-\fft{d^2\tilde R_r}{dr^2} + V_{\rm eff}^r \tilde R_r = (\omega^2)_r \tilde R_r + (V_{\rm eff}^i - (\omega^2)_i) \tilde R_i\,,\nn\\
&&-\fft{d^2\tilde R_i}{dr^2} + V_{\rm eff}^r \tilde R_i = (\omega^2)_r \tilde R_i - (V_{\rm eff}^i - (\omega^2)_i) \tilde R_r\,,
\eea
where $\tilde R_r + i \tilde R_i = \tilde R$. It is clear that for real $\omega$, both $R_r$ and $R_i$ independently satisfy the same equation.  They become coupled equations when $\omega_i$ is nonvanishing. Our numerical analysis indicates that $\omega_i/\omega_r < 10^{-6}$ for all the bound states.  Therefore, in the coupled equations, $R_r$ and $R_i$ provide perturbative sources for the same leading equation for $R_i$ and $R_r$ respectively. In this perturbative approach, the existence of trapping well can be determined by considering the real $\omega$ only.

\subsection{No superradiant instability of the RN black hole}\label{nogo}

Having determined the necessary conditions for the superradiant instability, we can search for bound state solutions of \eqref{schrlike}.  This in general requires numerical calculations.  However, in some simpler cases, we can determine from the effective potential that bound states are nonexisting. Such an example is provided by the RN black hole, which arises from the STU model by taking $\tilde Q_1=\tilde Q_2=\tilde Q_3=\tilde Q_4=Q/2$. As was argued previously, we consider the case with real $\omega$.  The bound state boundary condition and superradiant condition are
\begin{equation}\label{cond3}
\omega<m_p\,,\qquad \omega <\omega_c\,.
\end{equation}
To examine the shape of $V_{\rm eff}$, we define two positive real numbers $(x,y)$:
\begin{equation}\label{xy}
m_p=\omega (1 + x)\,,\qquad
\omega_c=\omega (1+y)\,,
\end{equation}
we find that the effective potential becomes
\begin{eqnarray}
V_{\rm eff} &=&(x+1)^2 \omega ^2 +\frac{4 Q \omega ^2 \left(x^2+2 x+y\right)}{r} +
\frac{\ell(\ell+1)+4 Q^2 \omega ^2 \left(x^2+2 x-(y-4) y\right)}{r^2}\nn\\
&&-\frac{16 Q^3 (y-1) y \omega ^2}{r^3}-\frac{16 Q^4 y^2 \omega ^2}{r^4}\,.
\end{eqnarray}
The bound state condition and superradiant condition require that
\begin{equation}
x > 0 \, , \qquad y > 0 \, .
\end{equation}
We see that the leading $1/r$ falloff at large $r$ is positive and the potential is negatives as $r\rightarrow 0$.  This implies that we must have at least three extrema at $r>0$
in order to have any bound states.  We find
\begin{eqnarray}
V'_{\rm eff} &=& \fft{-4  Q \omega ^2 \left(x^2+2 x+y\right)}{r^5} U\,,\nn\\
U &=& r^3 + a_2 r^2 + a_1 r + a_0\,,
\end{eqnarray}
where
\begin{eqnarray}
a_2 &=&\frac{\ell(\ell+1)+4 Q^2 \omega ^2 \left(x^2+2 x-(y-4) y\right)}{2 Q \omega ^2 \left(x^2+2 x+y\right)}\,,\nn\\
a_1 &=& -\frac{12 Q^2 (y-1) y}{x^2+2 x+y}\,,\qquad a_0= -\frac{16 Q^3 y^2}{x^2+2 x+y}\,.
\end{eqnarray}
It is clear that $U$ has three roots $(r_1,r_2,r_3)$. If these three roots are all positive, then we must have
\begin{equation}
a_0 <0\,,\qquad a_1>0\,,\qquad a_2<0\,.
\end{equation}
The first equality is automatically satisfied, but
\begin{equation}
\left\{
  \begin{array}{ll}
    a_1>0: &\quad \rightarrow\qquad y<1\,, \\
    a_2<0: &\quad \rightarrow\qquad y>4\,.
  \end{array}
\right.
\end{equation}
Thus the conditions for having three real positive roots cannot be satisfied for this case.

\begin{figure}[htpb]
  \centering
  \includegraphics[width=0.4\textwidth]{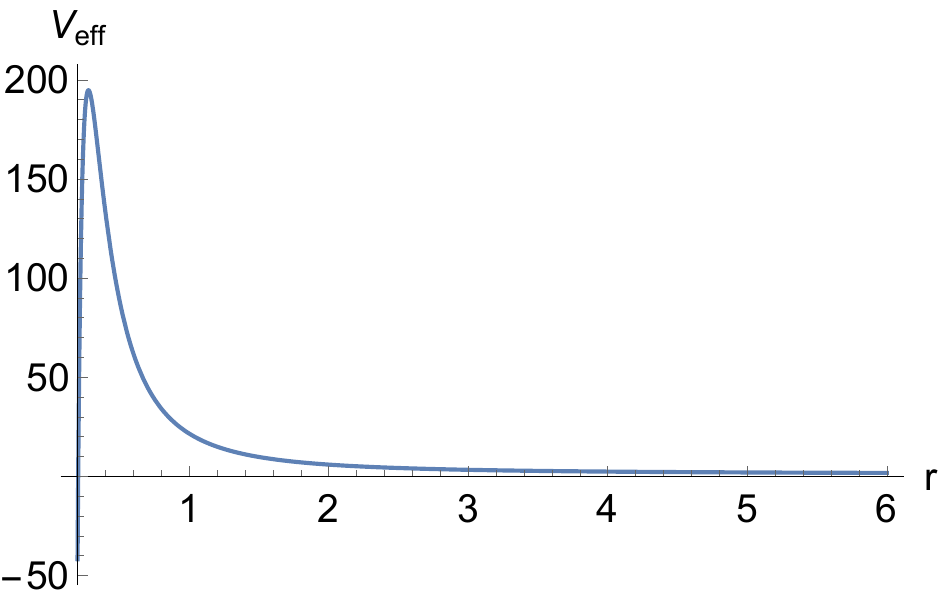}
 \caption{As an illustration, we plot the effective potential of the scalar radial wave in the extremal RN black hole for $\ell = 1$, $m_p = 1$, $\omega = 0.9$, $\tilde{q}_i =  \frac{1}{4}$ and  $Q =2$. This feature of having one extremal turns out to be generic when both the $\omega<m_p$ and $\omega <\omega_c$ conditions are satisfied.}\label{effRN}
\end{figure}

As an example, the associated effective potential $V_{\rm eff}$ for $\ell = 1$, $m_p = 1$, $\omega = 0.9$, $\tilde q_i = \frac{1}{4}$ and $Q = 2$ is plotted in Fig.~\ref{effRN}. It is obvious that even though the superradiant condition and bound state condition Eq.~\eqref{cond3} are satisfied, the superrdiant instability will not be triggered on since there is no trapping well for the effective potential.
Of course, the true bound states require that the frequency $\omega$ be complex, in which case, the discussion is far more complicated.  However, our numerical analysis indicates, for all the bound states associated with the STU charged black holes, that $\omega_i/\omega_r<10^{-6}$ and therefore the above conclusion is valid.

The absence of a trapping well in the effective potential illustrates that the interaction or the force between the charged scalar and the RN black hole is too ``simply'' structured involving only the gravitational attraction and charge repulsion. In order to create a trapping well, additional attractive force is necessarily outside the horizon of the black hole. This is one of the reasons why we study STU black holes which involve not only the Maxwell fields but also the dilatonic scalar fields that can give nontrivial interactions and therefore create possibly the needed trapping wells.

\section{Superradiant instability of STU black holes}\label{numer}

In the previous section, we show that in the RN black hole limit where all the charges are equal, there is no superradiant instability, regardless the configuration of the fundamental charges $\tilde q_i$.  In this section, we show that superradiance can develop instability when $\tilde Q_i$ are not all equal.  For simplicity, we shall consider the simplified two-charge solution  Eq.~\eqref{exbg}. In this case, the radial equation takes the same form of \eqref{radeq}, but with
\begin{equation}
U(r)=-r^4 H_1^{N_1} H_2^{N_2}\left(\omega-\sum_{i=1}^2 \fft{4\sqrt{N_i}\,q_i Q_i}{r H_i} \right)^2+m_p^4 r^4 H_1^{N_1/2}H_2^{N_2/2}+\ell(\ell+1)r^2 \, .
\end{equation}
In addition, the effective potential in the Schr\"odinger-like equation \eqref{schrlike} is given by
\begin{equation}\label{Veff}
V_{\rm eff} =\omega^2 + m_p^2 \sqrt{H_1^{N_1} H_2^{N_2}} -
\left(\omega-\sum_{i=1}^2 \fft{4\sqrt{N_i}\,q_i Q_i}{r H_i} \right)^2H_1^{N_1} H_2^{N_2}+ \fft{\ell(\ell+1)}{r^2}\,,
\end{equation}
where we require that the frequency $\omega$ satisfy both the bound state condition and superradiant condition
\begin{equation}
\omega<m_p\,,\qquad \omega<\omega_c =  \sqrt{N_1} q_1 + \sqrt{N_2} q_2 \,.
\end{equation}
In the RN black hole limit, $Q_1/\sqrt{N_1}=Q_2/\sqrt{N_2}$, it is straightforward to show that $V_{\rm eff}$ has no trapping well and hence there is no superradiant instability.

\subsection{Effective potentials with trapping well}

The effective potential with trapping well configuration does arise when the four charges in the STU model are not all equal. In this paper,  we shall focus on the $(N_1,N_2)=(2,2)$ and $(1,3)$ examples that are the special cases in STU supergravity. As was mentioned earlier, there are two types of trapping wells, shown in Fig.~\ref{veffshapes}, depending on whether the leading falloff parameter $\beta$ of the potential \eqref{vefffalloff} is positive or negative.
Here we present some concrete examples.

If we choose parameters
$(\omega, m_p, q_1,q_2,\ell,Q_1,Q_2)=$(18/100, 2/10, 2/10, 0, 1, 1, 8), the potential has two extrema for both $(N_1,N_2)=(2,2)$ and $(1,3)$ cases.  In the former case,  the peak $V_{\rm eff}^{\rm max}=2.93$ at $r=1.34$ and the trapping well minimum $V_{\rm eff}^{\rm min}=-0.0326$ at $r=7.27$.  In the latter case, we have $V_{\rm eff}^{\rm max}=51.99$ at $r=0.39$ and $V_{\rm eff}^{\rm min}=-6.14$ at $r=1.24$. The potentials all have the shape of the left panel in Fig.~\ref{veffshapes}.

If we choose the corresponding parameters as
(499/1000,5/10,2/10,2/10,1,10,1/10), the potential has instead three extrema: one trapping well minimum sandwiched between two peaks. For $(N_1,N_2)=(2,2)$ the peaks 685 and 0.26 are located at $r=0.07$ and 68.33, with the minimum -192 at 0.21.  For $(N_1,N_2)=(1,3)$, the minimum -0.11 at 1.51 is sandwiched between the two peaks 3983 and 0.29 located at 0.019 and 25.22 respectively. These potentials have the shape of the right panel in Fig.~\ref{veffshapes}.

While the existence of a trapping well potential is an encouraging sign, it does not guarantee that quasi-bound states must exist, since the existence depends on both the width and depth of the trapping well. We therefore adopt numerical approach next to search for the quasi-bound states in the parameter space where trapping wells arise.

\subsection{Numerical setup : the shooting method}

Here we briefly discuss our numerical approach for calculating the quasi-bound states. While the Schr\"odinger like equation \eqref{schrlike} guides us to find the parameter regions where solutions may exist, we actually use the equation \eqref{radeq} and its boundary condition Eq.~\eqref{asy1} to construct explicit numerical solutions. We adopt the shooting method to perform numerical calculations. As was shown in appendix \ref{qbmcom}, the frequency of these states must be complex $\omega=\omega_r + i \omega_i$ and the sign of $\omega_i$ plays a defining role whether a state is stable or unstable. However, the imaginary part of the frequency is generally small, compared to its real part. (See also the case of Kerr black hole in \cite{Dolan:2007mj}.) In order to avoid the cutoff error, accuracy issues must be taken care of in order to obtain the trustworthy results.

First we take the following asymptotic ansatz of the radial function as the boundary conditions near the horizon and asymptotic infinity:
\begin{equation}
R \left( r \to 0\right) \sim \left. \te^{i\frac{\chi_1 (\omega - \omega_c)}{r}}r^{-2i \chi_2}\sum^{n_1}_{i=0}r^i h_i \right|_{r = \epsilon_c } \,  ,  \quad
R \left( r \to \infty  \right) \sim  \left. \te^{- k r} r^{\chi_3} \sum^{n_2}_{i=0} \frac{g_i}{r^i}  \right|_{r = r_c}          \, ,
\end{equation}
where
\begin{eqnarray}
&& \chi_1 = 16 N_1^{-N_1/4}N_2^{-N_2/4}Q_1^{N_1/2}Q_2^{N_2/2} \, , \cr
&& \chi_2 = N_1^{-N_1 / 4} N_2^{-N_2/4}Q_1^{-1+N_1/2}Q_2^{-1+N_2/2} \Big(2N_2 Q_1 q_2 + 2 N_1 q_1 Q_2 \cr
&&\qquad+ (N_2^{3/2}Q_1 + N_1^{3/2}Q_2)(\omega-\omega_c) \Big) \, , \cr
&& \chi_3 =1-2 k (\sqrt{N_1}Q_1 + \sqrt{N_2}Q_2) + \frac{m_p^2 (\sqrt{N_1}Q_1 + \sqrt{N_2} Q_2 ) - 4(q_1 Q_1 + q_2 Q_2)\omega}{k} \, , \cr
&& k=\sqrt{m_p^2 - \omega^2} \, , \qquad \omega_c = \sqrt{N_1 }q_1 + \sqrt{N_2} q_2 \,  .
\end{eqnarray}
Since the equation is linear, we set $h_0=1=g_0$ without loss of generality, and we can obtain the coefficients $(h_i, g_i)$ analytically by solving the radial equation order by order. We present the explicit $(h_1,g_1)$ here
\begin{eqnarray}
h_1 &=&  \frac{i}{32(\omega - \omega_c)}\Bigg(16m_p^2 + \frac{16 (\ell(\ell+1)+2 \chi_2 (i + 2 \chi_2 ))}{\chi_1} -\frac{(N_1N_2)^{3/2}\chi_1}{Q_1 Q_2} \Bigg(\omega^2\cr
&&\left.- 2\omega\left(
q_2(N_2-1)\left(\frac{1}{\sqrt{N_2}}-\frac{Q_1}{N_1^{3/2}Q_2}\right) +
q_1(N_1-1)\left(\frac{1}{\sqrt{N_1}}-\frac{Q_2}{N_2^{3/2}Q_1}\right)
\right) \right.  \cr
&& \cr
&& +q_1^2 \left(\frac{\sqrt{N_1}Q_2(2N_1-3)}{N_2^{3/2}Q_1}-N_1 +2  \right) +  q_2^2 \left(\frac{\sqrt{N_2}Q_1(2N_2-3)}{N_1^{3/2}Q_2}-N_2 +2  \right) \cr
&&  +2 q_1 q_2 \left(\frac{(N_1 -1)Q_2}{N_2 Q_1}+ \frac{(N_2 -1)Q_1}{N_1 Q_2} - \frac{N_1 N_2 -3}{\sqrt{N_1N_2}}\right)\Bigg) \, , \cr
&& \cr
g_1 &=&  \frac{1}{2k}\left(\ell(\ell+1)- \chi_3 \left(1- 2k(\sqrt{N_1}Q_1 + \sqrt{N_2}Q_2 +\chi_3) \right) \right. \cr
&&- \frac{8(\sqrt{N_1}Q_2 - \sqrt{N_2}Q_1)(N_1 q_1 Q_2 - N_2 q_1 Q_1)\omega}{\sqrt{N_1N_2}}- \left(4(Q_1^2 + Q_2^2)(k^2-\omega^2) \right. \cr
&&+ \left. \left.(2(\sqrt{N_1}Q_1 + \sqrt{N_2}Q_2)\omega - 4(q_1 Q_1 + q_2 Q_2))^2 \right) \right) \, .
\end{eqnarray}
The expressions for higher order coefficients become increasingly complicated and we shall not give them here. These power series expansions are necessary since the radial equation is singular on the $r=0$ horizon and we cannot either take the $r=\infty$ boundary condition literally in the numerical approach. We denote $\epsilon_c$ and $r_c$ as the numerically cutoffs for the horizon and asymptotic infinity respectively. In general, the cutoff error can be numerically controlled under the maximum of $ {\cal O} \left(\epsilon_c^{n_1 + 1} \right)$ and $ {\cal O} \left( \frac{1}{r_c^{n_2 + 1}} \right) $ by the expansion power $(n_1, n_2)$. For the larger orders $(n_1,n_2)$ of the power expansions, we can use the smaller integration range $[\epsilon_c, r_c]$ for the desired accuracy.

Using Range-Kutta method, we integrate numerically from $\epsilon_c$ to some midpoint $r_i$, obtaining a numerical solution $R_1$; we also obtain the solution $R_2$ by integrating from $r_c$ to $r_i$. If $R_1$ and $R_2$ describes the same solution in different regions, then they should match at $r_i$. To match the solutions obtained from different regions, we require the Wronskian of $R_1, R_2$ at the middle point $r_i$ vanish, namely
\begin{equation}
W(R_1,R_2)= \left. \frac{R_1 R'_2 - R_2 R'_1}{|R_1| |R_2|} \right|_{r=r_i} = 0 \, .
\end{equation}
(It is important to note that the matching condition is invariant under $R_i \rightarrow c_i R_i$ for any nonvanishing constant $(c_1,c_2)$; therefore, we could have chosen $h_0=1=g_0$ without loss of generality.)  To implement this matching condition numerically, we consider an appropriate small dimensionless number $W_c$ such that we require
\be
r W(R_1,R_2)\Big|_{r=r_i} < W_c\,.\label{wcond}
\ee

Having established the shooting method for calculating the quasi-bound states, we can start by selecting a real $\omega$ such that the corresponding $V_{\rm eff}$ has a trapping well. We use a numerical finding root program to obtain the correct complex ${\omega_{\rm QBM}}$ that satisfies the numerical Wronskian condition \eqref{wcond}.  We then incrementally scan the $\omega$ in an appropriate range and obtain the spectrum of the quasi-bound states.  We can set $\omega_r$ positive without loss of generality, and we look for solutions with positive $\omega_i$ such that the wave function has an exponentially increasing factor $e^{\omega_i t}$, signaling superradiant instability.

Our numerical analysis indicates that the frequency ratio $\varpi\equiv\omega_i/\omega_r$ is smaller than $10^{-6}$, and therefore we need increase our numerical accuracy so that the results are trustworthy. However, the higher accuracy requirement when $\varpi$ is about $10^{-13}$ makes the calculation time consuming.  We thus set the cutoffs of the near horizon and the infinity to be $\epsilon_h \sim 10^{-2}$ and $r_c \sim 10^{2}$ respectively and take the expansion order $(n_1, n_2) = (10, 10)$. This leads to the cutoff errors under $10^{-21}$. Furthermore, we take the numerical Wronskian condition parameter $W_c\sim 10^{-16}$.  We believe that this ensures that $\varpi$ of the quasi-bound states can be numerically calculated within sufficient accuracy in the range $10^{-6}\sim 10^{-13}$. We carry out these calculation for both $(N_1,N_2)=(2,2)$ and $(1,3)$ cases and present the results next.

\subsection{Numerical Results}

In this subsection, we present our numerical results of quasi-bound states with superradiant instability. We set $\ell = 1$ for all our numerical calculation. It is useful to define a dimensionless parameter $\epsilon$ that measure the deviation of the STU black hole from the RN black hole:
\begin{equation}
\epsilon \equiv \frac{\sqrt{N_1}Q_2}{\sqrt{N_2 }Q_1} - 1\,.\label{epdef}
\end{equation}
The RN black hole corresponds to $\epsilon=0$, for which there is no superradiant instability.
The mass of the extremal black hole is then given by
\begin{equation}
M=\frac{Q_1}{\sqrt{N_1}}\left(4 + N_2 \epsilon \right).
\end{equation}
It is clear that $\epsilon$ lies in the region $-1<\epsilon<\infty$. The purpose of this paper is not to be exhaustive and we focus on the cases with $\epsilon>0$. (When $N_1=N_2=2$, the negative region of $\epsilon$ is equivalent to that of $\epsilon>0$.)

%
\subsubsection{Effective potential with single peak}

In this subsection, we present the partial spectrum of the quasi-bound states with superradiant instability, arising from the trapping-well potential with a single peak. In other words, we choose parameters such that the potential $V_{\rm eff}$ has the shape of the left panel in Fig.~\ref{veffshapes}. As in the case of the Schr\"odinger equation where the energy is quantised for bound states, the complex frequency of our classical quasi-bound states are also quantised. The characteristics of the quantisation can be described by the ``overtone'' number $n$, which counts the number of the peaks of the radial wave function $|R(r)|$.  The $n=1,2,\cdots,$ correspond to the ground state, first excited and second excited states, etc.

In Fig.~\ref{epN} of appendix \ref{app:pf}, we plot some explicit low-lying examples of $|R(r)|$ ($n=1,2,3)$ for various parameters. In all the plots we have fixed $m_p = \frac{2}{10}$, $q_1 = \frac{2}{10}$, $q_2=0$ and $Q_1=1$.  (Recall that we set $\ell=1$ for all our numerical calculation.) Note that in these plots we have normalized radial wave equation by
\begin{equation}
\int^{r_c}_{\epsilon_h} \sqrt{-\bar{g}}|R(r)|^2 = 1 \, ,
\end{equation}
where $\bar{g}$ denotes the determinant of the background metric. Although we present $|R|$ graphs for $n=1,2,3$, we actually obtained the results for all the low-lying examples up to and including $n=10$. The complex frequencies of these quasi-bound states are summarised in Table \ref{mp2} and \ref{mp3}.

\begin{table}[hbtp]
\centering
\begin{tabular}{c|c|c|c}
  \hline
  \hline
  $n$ & $\epsilon = 2.5$ & $\epsilon = 5$ & $ \epsilon = 8$ \\
  \hline
 $1$ & $0.19568 + 1.8307 \times 10^{-8} i$
     & $0.18108 + 4.0846 \times 10^{-7} i$
     & $0.16555 + 5.5423 \times 10^{-7} i$  \\
  \hline
  $2$& $0.19773 + 1.2333 \times 10^{-9} i$
     & $0.18741 + 5.9368 \times 10^{-7} i$
     & $0.17332 + 1.8124 \times 10^{-6} i$         \\
  \hline
 $3$& $0.19864 + 7.1132 \times 10^{-9} i$
    & $0.19144 + 5.5701 \times 10^{-7} i$
    & $0.17937 + 1.5825 \times 10^{-6} i$ \\
   \hline
 $4$& $0.19910 + 4.2111 \times 10^{-9} i$
    & $0.19400 + 4.4175 \times 10^{-7} i$
    & $0.18398 + 1.7047 \times 10^{-6} i$ \\
   \hline
 $5$ &$0.19937 + 2.6307 \times 10^{-9}i $
     &$0.19565 + 3.2839 \times 10^{-7}i $
     &$0.18748 + 1.6231 \times 10^{-6}i$ \\
   \hline
 $6$ &$0.19954 + 1.7313 \times 10^{-9} i$
     &$0.19674 + 2.3987 \times 10^{-7} i$
     &$0.19012 + 1.4352 \times 10^{-6} i$ \\
 \hline
 $7$ &$0.19964 + 1.1917 \times 10^{-9} i$
     &$0.19749 + 1.7591 \times 10^{-7} i$
     &$0.19211 + 1.2145 \times 10^{-6} i$ \\
  \hline
 $8$ & $0.19972 + 8.5168 \times 10^{-10} i$
     & $0.19801 + 1.3074 \times 10^{-7} i$
     & $0.19362 + 1.0027 \times 10^{-6} i$ \\
  \hline
 $9$ &$0.19977 + 6.2818 \times 10^{-10} i$
     &$0.19839 + 9.8786 \times 10^{-8} i$
     &$0.19577 + 8.1840 \times 10^{-7} i$ \\
  \hline
 $10$ &$0.19981 + 4.7578 \times 10^{-10} i$
     &$0.19868 + 7.5958 \times 10^{-8} i$
     &$0.19566 + 6.6543 \times 10^{-7} i$ \\

 \hline
 \hline
\end{tabular}
\caption{The explicit complex frequencies of low-lying overtone states with $m_p = \frac{2}{10}$, $q_1 = \frac{2}{10}$, $q_2 = 0$ and $Q_1=1$  in the $(N_1,N_2)=(2,2)$ case.}\label{mp2}
\end{table}

\begin{table}[hbtp]
\centering
\begin{tabular}{c|c|c|c}
  \hline
  \hline
  $n$ & $\epsilon = 2$ & $\epsilon =2.5$ & $ \epsilon = 3.5$ \\
  \hline
 $1$ & $0.17028 + 5.3213 \times 10^{-7} i$
     & $0.16253 + 5.9330 \times 10^{-7} i$
     & $0.15031 + 4.4746 \times 10^{-7} i$  \\
  \hline
  $2$& $0.17769 + 1.0241 \times 10^{-6} i$
     & $0.17036 + 1.2805 \times 10^{-6} i$
     & $0.15810 + 1.1444 \times 10^{-6} i$         \\
  \hline
 $3$& $0.18320 + 1.2619 \times 10^{-6} i$
    & $0.17650 + 1.7653 \times 10^{-6} i$
    & $0.16460 + 1.8423 \times 10^{-6} i$ \\
   \hline
 $4$& $0.18728 + 1.2735 \times 10^{-6} i$
    & $0.18131 + 1.9846 \times 10^{-6} i$
    & $0.17005 + 2.3954 \times 10^{-6} i$ \\
   \hline
 $5$ &$0.19028 + 1.1522 \times 10^{-6}i $
     &$0.18508 + 1.9873 \times 10^{-6}i $
     &$0.17462 + 2.7524 \times 10^{-6}i$ \\
   \hline
 $6$ &$0.19250 + 9.7835 \times 10^{-7} i$
     &$0.18804 + 1.8513 \times 10^{-6} i$
     &$0.17846 + 2.9196 \times 10^{-6} i$ \\
 \hline
 $7$ &$0.19415 + 8.0070 \times 10^{-7} i$
     &$0.19034 + 1.6454 \times 10^{-6} i$
     &$0.18168 + 2.9307 \times 10^{-6} i$ \\
  \hline
 $8$ & $0.19537 + 6.4235 \times 10^{-7} i$
     & $0.19215 + 1.4182 \times 10^{-6} i$
     & $0.18439 + 2.8276 \times 10^{-6} i$ \\
  \hline
 $9$ &$0.19629 + 5.1069 \times 10^{-7} i$
     &$0.19357 + 1.1985 \times 10^{-6} i$
     &$0.18666 + 2.6498 \times 10^{-6} i$ \\
  \hline
 $10$&$0.19699 + 4.0521 \times 10^{-7} i$
     &$0.19469 + 1.0009 \times 10^{-6} i$
     &$0.18856 + 2.4297 \times 10^{-6} i$ \\

 \hline
 \hline
\end{tabular}
\caption{The explicit complex frequencies of low-lying overtone states with $m_p = \frac{2}{10}$, $q_1 = \frac{2}{10}$, $q_2 = 0$ and $Q_1=1$  in the $(N_1,N_2)=(1,3)$ case.}\label{mp3}
\end{table}

In a quantum mechanical system, $\omega_r$ would be the quantized energy of the system and the overtone number $n$ would describe the energy level. For our classical quasi-bound states, $\omega_r$ is also a monotonously increasing function of $n$ with the ground state ($n=1$) having the lowest frequency. On the other hand, $\omega_i$ does not change monotonously as we increase $n$, indicating that the most unstable mode in general is neither the ground state, nor the most excited state. Intriguingly, we find that the $n$-dependence of complex frequency can be accurately fitted with some simple close form functions:
\begin{eqnarray}\label{fitf1}
\frac{\omega_r}{M} = \frac{1+a_1 n}{1+b_1 n } \omega_{r}^0 \, , \qquad \frac{\omega_i}{M} = \frac{1+a_2 n^2}{1+b_2 n^2 +c_2 n^4}\omega_i^0 \, , \quad (n=1,2,3,\ldots)\, .
\end{eqnarray}
The coefficients $(\omega_{r_0}, a_1, b_1, \omega_{i_0}, a_2, b_2)$ depend on all the variables of the theory and the charge parameters of the solution. For the data set in Tables \ref{mp2} and \ref{mp3}, we give these coefficients in Table \ref{fitt1}.
\begin{table}[htpb]
\centering
\begin{tabular}{c|c|c|c|c|c|c|c|c}
  \hline
  \hline
  $(N_1, N_2$)&$\epsilon$ & $\omega_r^0 $ & $a_1 $ & $ b_1 $ & $\omega_i^0$ & $a_2$  & $b_2$ & $c_2$ \\
  \hline
 \multirow{4}{*}[7pt]{$(2,2)$}  &   $2.5$ & $0.0275$
                                          & $4.87$
                                          & $4.26$
                                          & $3.31\times 10^{-9}$
                                          & $0.0328$
                                          & $0.174$
                                          & $0.0153$\\
 \cline{2-9}
            ~             & $5$           & $0.0165$
                                          & $0.977$
                                          & $0.783$
                                          & $3.00\times 10^{-8}$
                                          & $0.484$
                                          & $0.0577$
                                          & $0.0157$\\
  \cline{2-9}
            ~             &  $8$         & $0.0107$
                                          & $0.417$
                                          & $0.299$
                                          & $2.17\times 10^{-8}$
                                          & $0.984$
                                          & $0.0667$
                                          & $0.00352$\\

 \hline
 \multirow{4}{*}[7pt]{$(1,3)$}  &  $2$ & $0.0156$
                                          & $0.476$
                                          & $0.356$
                                          & $3.23 \times 10^{-8}$
                                          & $0.814$
                                          & $0.0753$
                                          & $0.00510$\\
 \cline{2-9}
            ~                   & $2.5$  & $0.0130$
                                          & $0.348$
                                          & $0.243$
                                          & $2.81\times 10^{-8}$
                                          & $1.03$
                                          & $0.0756$
                                          & $0.00236$\\
  \cline{2-9}
            ~                  &  $3.5$  & $0.00963$
                                          & $0.230$
                                          & $0.144$
                                          & $1.41\times 10^{-8}$
                                          & $1.40$
                                          & $0.0530$
                                          & $0.000548$\\

\hline
\hline
\end{tabular}
\caption{The explicit coefficients of the fitting functions in Eq.~\eqref{fitf1}.}\label{fitt1}
\end{table}

In appendix \ref{app:pf}, we present Fig.~\ref{22vN1} and Fig.~\ref{22vN2}, where we plot the dimensionless real and imaginary frequencies $\omega_r/M$ and $\omega_i/M$ with respect to the overtone number $n$ for various $\epsilon$'s. The dots are the actual numerical results for the low-lying overtone numbers and the solid lines are produced by the fitting functions.
As we can see from Fig.~\ref{22vN1} and Fig.~\ref{22vN2}, our conjectured formulae in \eqref{fitf1} match with the quantized frequency strikingly well. The results show that $\omega_r(n)$ increases monotonously while $\omega_i(n)$ can have a maximum. The results also
indicate that for large $n$, we have
\be
\omega_r-\omega_r^0\sim - \fft{1}{n}\,,\qquad \omega_i \sim \fft{\omega_i^0}{n^2}\,.
\ee
Thus the instability are not caused by modes with large overtones.

We now turn to the discussion of how $\omega_i$ depends on the charge configuration, described by the dimensionless parameter $\epsilon$. This is important since we have already seen that there is no quasi-bound state when $\epsilon=0$.  We thus naturally expect that quasi-bound states would emerge only for sufficiently large $\epsilon$. We focus our discussion for overtone numbers $n=1$ and $n=2$.  We also fixed parameters $m_p=\frac{2}{10}$ and $q_1=\frac{2}{10}$ and $q_2=0$.  We present our results in Fig.~\ref{epome} of appendix \ref{app:pf}. To better describe the $\epsilon$ dependence, we find that the numerical data can be fitted by an exponential function, given by
\begin{equation}\label{fitf2}
 \log \frac{\omega_i}{M} = a_3 \epsilon^{-\frac{N_1 + N_2}{2}}+ b_3 \epsilon^{-\frac{N_2}{2}} + c_3 \epsilon^{-\frac{N_1}{2}} + d_3 + e_3 \epsilon^{\frac{N_1}{2}} + f_3 \epsilon^{\frac{N_2}{2}}  \, .
\end{equation}
In the case of $(N_1, N_3)=(1,3)$, the values of the coefficients $(a_3, b_3, c_3, d_3, e_3, f_3)$ are given in Table \ref{fitt2}.
\begin{table}[hbtp]
\centering
\begin{tabular}{c|c|c|c|c|c|c}
  \hline
  \hline
 $n$ & $a_3 $ & $b_3 $ & $ c_3 $ & $d_3$ & $e_3$ & $f_3$\\
  \hline
                                 $1$ & $8.43$
                                     & $-21.5$
                                     & $9.62$
                                     & $-9.65$
                                     & $-6.49$
                                     & $0.295$  \\
\hline
                                 $2$ & $21.5$
                                      & $-59.0$
                                      & $93.5$
                                      & $-99.3$
                                      & $25.1$
                                      & $-1.04$  \\

 \hline
 \hline
\end{tabular}
\caption{The value of $(a_3, b_3, c_3, d_3, e_3, f_3)$ for the fitting function Eq.~\eqref{fitf2} in the $(N_1, N_2)=(1, 3)$ case.}\label{fitt2}
\end{table}

For the case of $(N_1, N_2)=(2,2)$, the Eq.~\eqref{fitf2} reduces to a simpler function:
\begin{equation}\label{fitf22}
 \log \frac{\omega_i}{M}=  \tilde a_3 \epsilon^{-2} + \tilde b_3 \epsilon^{-1} + \tilde c_3 + \tilde d_3 \epsilon \, .
\end{equation}
The coefficients $(\tilde a_3, \tilde b_3, \tilde c_3 , \tilde d_3)$ are given in Table \ref{fitt22}.
\begin{table}[htbp]
\centering
\begin{tabular}{c|c|c|c|c}
\hline
\hline
 $n$ & $\tilde a_3 $ & $\tilde b_3$ &$\tilde c_3 $ &    $\tilde d_3$ \\
  \hline
$1$ & $-1.97$
    & $-16.9$
    & $-11.5$
    & $0.422$   \\
\hline
$2$ & $2.09$
    & $-22.9$
    & $-10.2$
    & $-0.403$  \\

 \hline
 \hline
\end{tabular}
\caption{The value of coefficients $(\tilde a_3, \tilde b_3, \tilde c_3 , \tilde d_3)$ for the fitting function Eq.~\eqref{fitf22} in the $(N_1, N_2)=(2,2)$ case.}\label{fitt22}
\end{table}

Our numerical data of $\omega_i$ run from $10^{-12}$ to $10^{-6}$ as $\epsilon$ increases from small value to a larger one, and that is quite a few orders of magnitude. For this reason, we plot our results in Fig.~\ref{epome}, using logarithmic $\log (\omega_i/M)$. The plots illustrate that our conjectured curve functions fit with the data remarkably well.

We now examine the example of $(N_1,N_2)=(2,2)$ and $n=1$ in further detail. We have computed quasi-bound states for parameter $\epsilon$ from about 1 to 12. We see that the imaginary part of the dimensionless frequency has a maximum of $\sim 10^{-6}$ around $\epsilon\sim 6$, and it becomes vanishingly small ($\epsilon\sim 10^{-12}$) as $\epsilon\sim 1$.  This appears to have confirmed our earlier expectation that quasi-bound states cease to exist for sufficiently small $\epsilon$.  This feature remains true for the $n=2$ states, or for the $(N_1,N_2)=(1,3)$ theory.

However, there is a caveat in our above discussion, since we set $q_2=0$. We might naively conclude that the black holes with small enough $\epsilon$ might be exempted from superradiant instability. This obviously appears to be consistent with the fact that the RN black hole $(\epsilon=0)$ is stable. However, the situation changes if we turn on $q_2$ parameter.  To illustrate this, we consider the $(N_1,N_2)=(2,2)$ example and the leading falloff of the large $r$ expansion of the effective potential is
\begin{equation}\label{Veff2}
V_{\rm eff} = m_p^2 + \fft{\beta}{r} + {\cal O}(1/r^2)\,,
\end{equation}
where
\begin{eqnarray}
\beta &=& 2Q_1\Big(2 \sqrt{2} (x+y \omega ) + \epsilon(
\sqrt2 (x-\omega^2) + 4q_2\omega)\Big),\nn\\
x&=& m_p^2 - \omega^2>0\,,\qquad y=\sqrt2(q_1+q_2)-\omega>0\,.\label{betaeq}
\end{eqnarray}
As was explained earlier, for $V_{\rm eff}$ to have one peak, we must have $\beta<0$.  This is not possible for the RN black hole with $\epsilon=0$. When $q_2=0$, we find that $\beta$ can only be negative for sufficiently large $\epsilon$ since $\omega <m_p$.  However, if we turn on $q_2$ and let it be negative, then we can have negative $\beta$ even for smaller $\epsilon$.
Of course, the condition $y>0$ restrict how negative $q_2$ can go.  Analogous results can be found for the $(N_1,N_2)=(1,3)$ theory.

It becomes clear why trapping wells can exist in STU black holes while they do not in the RN black hole.  Having negative $\beta$ is a sign that the overall force becomes attractive at the asymptotic infinity, and this requirement can be arranged in the STU theory since we can have additional antiparticles with negative $q_2$ that is attractive when it interacts with $Q_2$. These attractive components, mediated also through the dilaton scalar, create an effective potential with a trapping well. This also implies that superradiant instability could exist as long as the STU black hole has not degenerate to become the RN black hole and hence the dilatonic scalars are not vanishing.

However, having found the condition for negative $\beta$ provides only the necessary condition for the quasi-bound states and we still resort to numerical calculations for more quasi-bound states. We start from the least unstable state among those in Fig.~\ref{epome}, namely $\epsilon=1$ and $q_2=0$ for which we have seen that $\omega_i/M\sim 10^{-14}$ is vanishingly small. Fig.~\ref{epome} appears to suggest that black holes with smaller $\epsilon$, e.g.~$\epsilon=0.5$, would become stable. However if we fix the parameters $m_p = \frac{2}{10}$, $q_1 = \frac{2}{10}$, $Q_1=1$ and $\epsilon=1$, but turn on  $q_2$ so that $q_2/q_1$ runs from 0 to some small negative number, we find that $\omega_i/M$ dramatically increases by a few orders of magnitudes. Explicitly, from the panels of the first row in Fig.~\ref{q2ome} of appendix \ref{app:pf}, in which we plot the $\omega_i/M$ in terms of the fundamental charge ratio of $q_2/q_1$. We see that when this ratio is zero, $\omega_i/M$  is vanishingly small, around $10^{-14}$. However, it increases from $10^{-14}$ to around the maximum $10^{-10}$ and then decreases again, as we make the ratio more and more negative.

We therefore select $q_2 = -0.048$, corresponding to the maximum of $\omega_i / M \approx 3.8 \times 10^{-10}$, and fix also parameters $m_p = \frac{2}{10}$, $q_1 = \frac{2}{10}$, $Q_1=1$.
We construct more quasi-bound states by decreasing $\epsilon$ from 1 to some smaller number. We find that the instability modes exist for smaller $\epsilon$, with $\omega_i/M $decreasing to to $10^{-13}$ as $\epsilon$ runs from $1$ to around $0.3$. Similarly, in the case of $q_2 = -0.048$, the fitting functions in Eq.~\eqref{fitf22} for $\omega_i/M$ with respect to $\epsilon$ is also valid, and the panels of the second row of Fig.~\ref{q2ome} show that they fit remarkably well. The coefficients $(\tilde a_3, \tilde b_3, \tilde c_3 , \tilde d_3)$ are explicitly given in Table \ref{fitt22v2}.
\begin{table}[htbp]
\centering
\begin{tabular}{c|c|c|c|c}
\hline
\hline
 $n$ & $\tilde a_3 $ & $\tilde b_3$ &$\tilde c_3 $ &    $\tilde d_3$ \\
  \hline
$1$ & $0.247$
    & $-3.35$
    & $-24.1$
    & $5.53$   \\
\hline
$2$ & $0.1819$
    & $-2.92$
    & $-26.1$
    & $6.33$  \\

 \hline
 \hline
\end{tabular}
\caption{The value of coefficients $(\tilde a_3, \tilde b_3, \tilde c_3 , \tilde d_3)$ for the fitting function Eq.~\eqref{fitf22} in $(N_1, N_2)=(2,2), q_2 = -0.048$ case.}\label{fitt22v2}
\end{table}

The above results strongly suggest that STU black holes with nonvanishing $\epsilon$ is always unstable. Indeed, from \eqref{betaeq}, we can see that no matter how small $\epsilon\ne 0$ is, we can always increase $q_1$ positively and make $q_2$ more negative so that $\beta$ becomes negative while keeping $q_1 + q_2$ fixed to satisfy the superradiant condition. Concretely, we obtain unstable quasi-bound states for $Q_1=1$, $\epsilon=0.01$, with the complex frequency
\begin{equation}
\omega=0.1927 + 3.752\times 10^{-8}i\,,
\end{equation}
when we choose $m_p = \frac{2}{10}$, $q_1 = 20.2$, $q_2=-20$.  We may therefore conclude that the superradiant stability of the RN black hole is a fine-tuning result from the perspective of the STU supergravity model.

\subsubsection{Effective potential with double peaks}

As was discussed earlier, at large $r$, the leading falloff of $V_{\rm eff}$ takes the form
of \eqref{Veff2}.  In the previous subsection, we consider the case with $\beta <0$, in which case the potential has a single peak.  Here, we consider the possibility of $\beta>0$, in which case, a potential trapping well necessarily involves two peaks. For simplicity, we only investigate the $(N_1, N_2)=(2,2)$ example. A simple way to achieve $\beta>0$ is to take $q_2=q_1$, in which case, we have
\be
\beta = 2 \sqrt{2} \left(Q_1+Q_2\right) \omega ^2 \left(x^2+2 x+y\right),
\ee
where $x>0$ and $y>0$ are defined by \eqref{xy} with $\omega_c=2\sqrt2 q_1$. As we have shown, for the RN black hole with $Q_2=Q_1$, the potential $V_{\rm eff}$ shapes like Fig.~\ref{effRN} and therefore has no trapping well. It turns out that if we take an alternative limit with $Q_1=0$ (or equivalently $Q_2=0$), there is no trapping well either.  However, trapping well does arise when we keep $Q_2/Q_1$ sufficiently large, but not too large, corresponding to a large but not too large $\epsilon$, defined by \eqref{epdef}.

In what follows, we illustrate the superradiant instability associated with quasi-bound states in the trapping-well potential with double peaks. We fix parameters $m_p = \frac{5}{100}$, $ q_1 = q_2 = \frac{2}{100}$, $Q_2 = 100$, and let $Q_1$ to run from $1$ to $0.04$, corresponding to $\epsilon$ running from $99$ to $2499$.

We begin with presenting the quasi-bound states with fixed $Q_1=1$ ($\epsilon=99$) and $Q_1=0.1$ ($\epsilon=999$) for some low-lying overtone number $n$. We find that the complex frequency as a function of $n$ can also be fitted with our conjectured formulae in Eq.~\eqref{fitf1}, with coefficients $(\omega_{r_0}, a_1, b_1, \omega_{i_0}, a_2, b_2)$ given in Table \ref{fittD1}.
As we can see from Fig.~\ref{22vDN1}, the solid lines drawn by the fitting functions match with the numerical data quite well.
\begin{table}[htpb]
\centering
\begin{tabular}{c|c|c|c|c|c|c|c}
  \hline
  \hline
 $\epsilon$ & $\omega_r^0 $ & $a_1 $ & $ b_1 $ & $\omega_i^0$ & $a_2$  & $b_2$ & $c_2$ \\
  \hline
                           $99$           & $2.90\times 10^{-4}$
                                          & $0.147$
                                          & $0.101$
                                          & $2.49\times 10^{-9}$
                                          & $0.76$
                                          & $0.0271$
                                          & $0.000807$\\
\hline

                                    $999$ & $2.58 \times 10^{-4}$
                                          & $0.0794$
                                          & $0.0344$
                                          & $1.59\times 10^{-8}$
                                          & $0.621$
                                          & $0.0704$
                                          & $0.000466$\\
\hline
\hline
\end{tabular}
\caption{The explicit $(\omega_{r_0}, a_1, b_1, \omega_{i_0}, a_2, b_2)$ for Eq.~\eqref{fitf1} in the $(N_1, N_2) = (2,2)$ case, with $m_p = \frac{5}{100}$, $ q_1 = q_2 = \frac{2}{100}$, $Q_2 = 100$. The corresponding $\omega_r(n)$ and $\omega_i(n)$ produce the solid lines in Fig.~\ref{22vDN1}.}\label{fittD1}
\end{table}

Finally, we turn to investigate how the dimensionless instability parameter $\omega_i/M$ changes according to the value of $\epsilon$. Similarly, we study only the ground state ($n=1$) and the first excited state ($n=2$) as examples. As was mentioned, we expect the quasi-bound states do not exist for small $\epsilon$ and neither for $\epsilon$ that is too large.
In Fig.~\ref{epomeD}, we present $\log(\omega_i / M)$ as functions of $\epsilon$ that runs from 99 to 2499. Even though we have obtained a large region of $\epsilon$, we find that our conjectured formula Eq.~\eqref{fitf22} with coefficients given in Table \ref{fitt22v3} can fit the data perfectly well.
\begin{table}[htbp]
\centering
\begin{tabular}{c|c|c|c|c}
\hline
\hline
 $n$ & $\tilde a_3 $ & $\tilde b_3$ &$\tilde c_3 $ &    $\tilde d_3$ \\
  \hline
                                 $1$ & $-197$
                                     & $-184$
                                     & $-17.3$
                                     & $-0.000116$   \\
\hline
                                $2$ & $132$
                                     & $-203$
                                     & $-16.5$
                                     & $-0.000163$  \\

 \hline
 \hline
\end{tabular}
\caption{The value of coefficients $(\tilde a_3, \tilde b_3, \tilde c_3 , \tilde d_3)$ for the fitting function Eq.~\eqref{fitf22} in the $(N_1, N_2)=(2,2)$ case,
with $m_p = \frac{5}{100}$, $ q_1 = q_2 = \frac{2}{100}$, $Q_2 = 100$.}\label{fitt22v3}
\end{table}

\section{Conclusions}\label{conclu}

In this paper, we studied superradiant instability by constructing quasi-bound states of charged Klein-Gordon scalar equation in the background of extremal charged black holes of the STU supergravity model. The general black hole carries four $U(1)$ charges and when they are all equal, it reduces to the RN black hole. We were motivated from the fact that the RN black hole seems to be exempted from the superradiant instability and we would like to study this phenomenon from the perspective of these more general charged black holes.

We first addressed some theoretical aspects of the subject. We proved that all quasi-bound states, as well as the quasinormal modes, necessarily involve complex frequencies. This motivated us to establish superradiant boundary conditions for such states with complex frequencies. Owing to the extra $U(1)$ fields and scalars in the STU model, the effective potential of the Schr\"odinger-like equation governing the radial wave function has far richer structure than the one associated with the RN black hole. For simplicity, we reduced the general STU model to involve only one scalar and two $U(1)$ fields. The reduced system with the nontrivial dilaton captures the essence of the STU model.

We found that for suitable parameters, there exist two types of trapping wells: those with single peak and those with double peaks. Using the numerical shooting method, we found that both potential wells could trap quasi-bound states with positive imaginary frequency, indicating superradiant instability. These are the first examples in literature where superradiant instability is demonstrated for spherically-symmetric and static black holes. We obtained how the complex frequency of the quasi-bound states depended on the overtone number, as well as the parameter $\epsilon$ that measures the deviation of the STU black holes from the RN black hole.

Our results indicate that STU charged black holes are superradiant unstable as long as the charges are not equal. In other words, no matter how small $\epsilon$ is, we can always find superradiant instability associated with appropriate fundamental charged massive scalar particles and antiparticles. Therefore the superradiant stability of the RN black hole is a fine-tuning result in the framework of the STU supergravity theory.

It is clear that the electric repulsion and gravity attraction associated with RN black hole and the superradiant particle is too simple for the effective potential to have a potential well. The scalar fields in the STU model can yield both attractive and repulsive long ranged forces, and this makes it possible for creating a potential trapping well. It is thus of great interest to study generally how scalar fields affect the superradiant (in)stability of
black holes with the scalar hair.

\section{Acknowledgement}

We thank Bing Sun for useful discussions. This work is supported by the National Natural Science Foundation of China (NSFC) grants No.~12005155, No.~11875200 and No.~11935009, and also No.~11947301 and No.~12047502.

\section*{Appendix}

\appendix

\section{Tortoise Coordinate}\label{tort}

To investigate the asymptotic behavior of quasi-bound states, it is useful to introduce the following definition of tortoise coordinate $y$ and radial function transformation $\bar{R}$

\begin{equation}\label{barR}
 \frac{\td y}{ \td r} = (\tilde H_1 \tilde H_2 \tilde H_3 \tilde H_4)^{\frac{1}{2}}\, ,    \qquad  \bar{R}  \equiv   \left(\tilde H_1 \tilde H_2 \tilde H_3 \tilde H_4\right)^{-1/4}\frac{R(r)}{r}\, ,
\end{equation}
where $y \to -\infty $ as $r \to 0$ while $y \to \infty$ as $r \to \infty$. Then the radial equation can be written in standard wave equation
\begin{equation}\label{radeq3}
-\frac{\td^2 \bar{R}}{\td y^2} + \bar U(y)\, \bar{R} = 0 \,,\qquad \bar U=-(\omega - Q(y))^2 + V(y)\,,
\end{equation}
where
\begin{equation}
Q(y)=\sum_{i=1}^4 q_i \left( 1- \tilde H_i^{-1}   \right) \, , \quad V(y)= \frac{\ell (\ell + 1)}{r^2 \tilde{H}}+ \frac{m_p^2}{\sqrt{\tilde H}} - \frac{5 \tilde H'}{16 \tilde H^3} + \frac{\tilde H''}{4 \tilde H^2} \, , \quad \tilde H = \prod^4_{i=1}\tilde H_i
\end{equation}
The potential $\bar{U}$ has the following asymptotic behavior
\begin{equation}\label{Uasy}
\bar{U}|_{ y \to -\infty}  \sim -(\omega - \omega_c)^2  \,  , \quad \bar{U}_{y \to \infty}  = m_p^2 - \omega^2>0 \, .
\end{equation}
With such asymptotic behavior of $\tilde{U}$, we will arrive the asymptotic solution of $\tilde{R}$ that
\begin{equation}\label{asyso2}
\bar{R} \varpropto \left\{
\begin{aligned}
&\te^{-i (\omega - \omega_c)y}  \quad \quad  &&y \to -\infty \\
& \te^{- \sqrt{m_p^2 - \omega^2 }y} \quad \quad && y \to \infty \, ,
\end{aligned}
\right.
\end{equation}
where we discarded the divergent mode and the wave function decays exponentially at infinity, giving rise to a quasi-bound state. Compared to the original $r$ variable, in the two asymptotic regions, we have
\begin{equation}\label{yr}
y \sim - \frac{1}{r} \, ,\quad \bar {R} \sim R \, ,  \quad \text{as} \quad r \to 0\, ; \qquad y \sim r \, , \quad \bar{R} \sim \frac{1}{r} R  \, , \quad \text {as} \quad r \to \infty \, . \end{equation}
Eq.~\eqref{asyso2} can then be translated to the boundary conditions in $r$ coordinate, namely Eq.~\eqref{asy1}. Thus we see that $r=0$ horizon with ingoing boundary condition is not an infinite potential barrier, but a non-repulsive asymptotic region.

It is worth remarking that when $\omega>m_p$, the function $\bar R$ becomes wavelike also as $y\rightarrow \infty$. A quasinormal mode is defined as the one with outgoing boundary condition only, namely $\bar R\sim \te^{i \sqrt{\omega^2-m_p^2}\,y}$ as $y\rightarrow \infty$.

\section{A proof of complex frequencies}\label{qbmcom}
\begin{theorem}\label{KNreal}
The time frequency $\omega$ of quasinormal modes or quasi-bound states cannot be a real number.
\end{theorem}
\textit{Proof}:  We begin by assuming that the frequency of quasinormal modes or quasi-bound states is a real number. By performing integration by parts on the radial equation Eq.~\eqref{radeq3}, we have
\begin{equation}\label{RddR1}
\left. \bar R_n^{*}\frac{\td \bar R_n}{\td y} \right|^{\epsilon}_{-\epsilon}=\int^{\epsilon}_{-\epsilon} \left(|\bar R_n'|^2 -\left((\omega_n-Q(y))^2-V(y) \right)|\bar R_n|^2 \right) \td y \, ,
\end{equation}
where a prime denotes a derivative with respect to $y$. We set $\epsilon \gg 1$, but not infinity, to avoid divergence. With boundary conditions that define either the quasinormal modes or quasi-bound states, the left-hand side of Eq.~\eqref{RddR1} gives
\begin{eqnarray}
&& {\rm quasinormal~modes:} \quad \left. \bar R_n^{*}\frac{\td \bar R_n}{\td y} \right|^{\epsilon}_{-\epsilon} = i \left(|c_1|^2(\omega-\omega_c)+|c_2|^2\sqrt{\omega^2-m_p^2}\right) \, , \quad  m_p^2<\omega^2  \cr
~\cr
&& {\rm quasi-bound states:} \quad \left. \bar R_n^{*}\frac{\td \bar R_n}{\td y} \right|^{\epsilon}_{-\epsilon} = i |c_1|^2(\omega-\omega_c) \, , \quad m_p^2 > \omega^2 \,,
\end{eqnarray}
which is purely imaginary.  On the other hand, the right-hand side of Eq.~\eqref{RddR1} must be real, since $\omega$, $Q(y)$ and $V(y)$ are all real, i.e.
\begin{equation}\label{RddR2R}
\int^{\epsilon}_{-\epsilon} \left(|\bar R_n'|^2 -\left((\omega_n-Q(y))^2-V(y) \right)|\bar R_n|^2 \right) \td y \in \mathbb{R} \, .
\end{equation}
The contradiction shows that neither the quasinormal modes nor quasi-bound states can have a real time frequency. It should be pointed that our proof relies only on the boundary conditions of either quasinormal modes or quasi-bound states. It does not depend on the details such as $V(y)$ and $Q(y)$, and therefore the statement is completely general.

\section{Plots and figures}
\label{app:pf}

In this appendix, we give our plots and figures discussed in section \ref{numer}.  The explanations of graphs are provided by the captions and also the associated discussions in the main text.

\begin{figure}[htpb]
  \centering
  \includegraphics[width=0.32\textwidth]{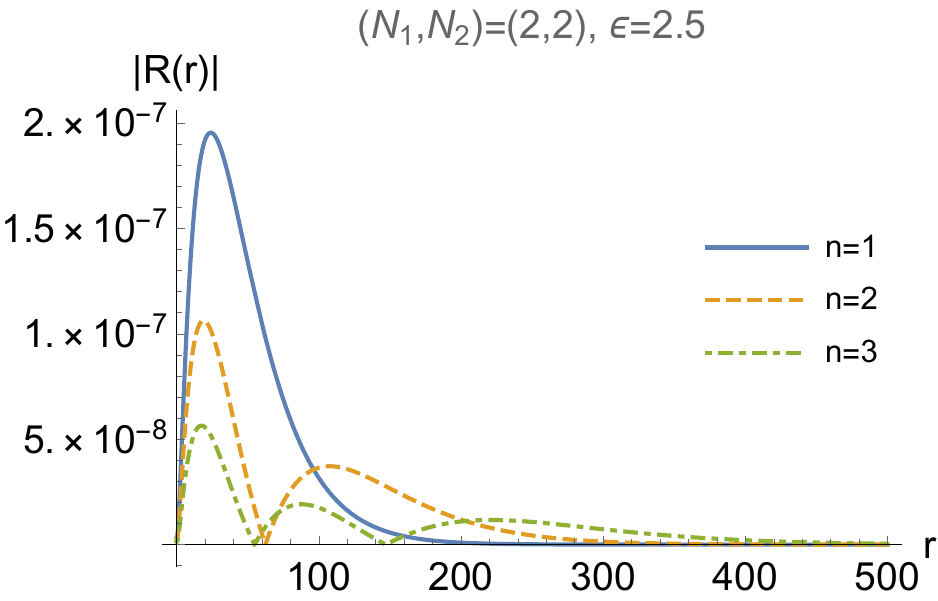}
  \includegraphics[width=0.32\textwidth]{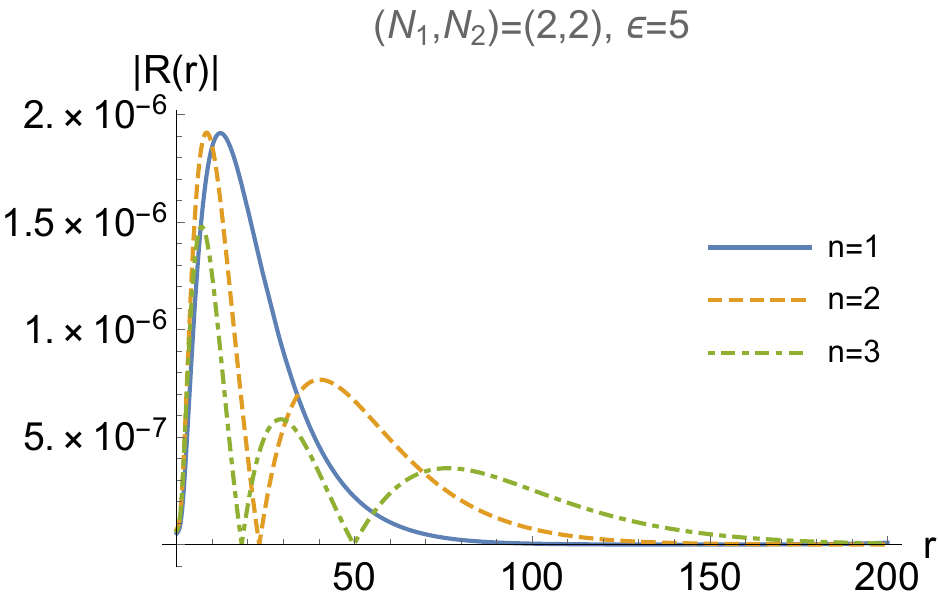}
  \includegraphics[width=0.32\textwidth]{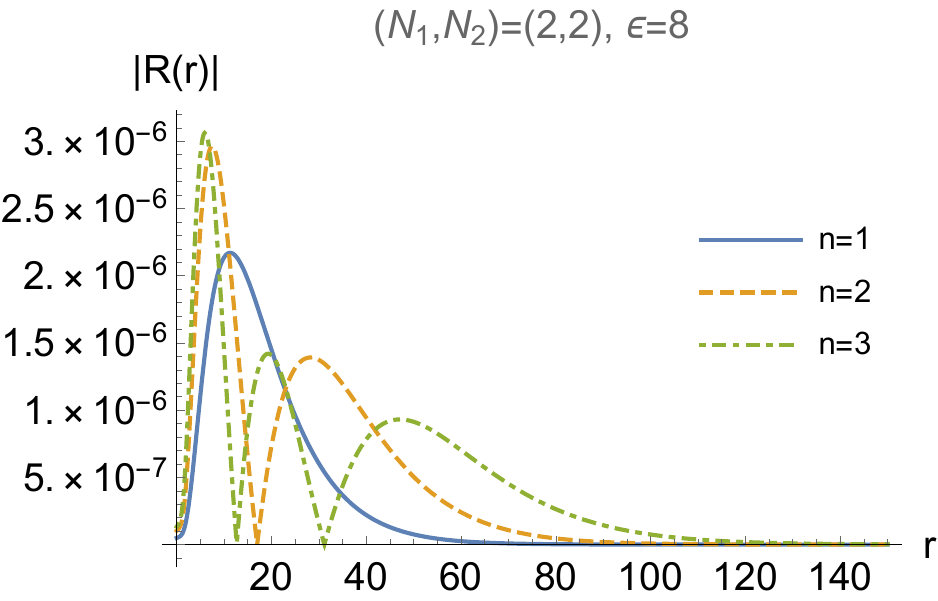}\\

  \includegraphics[width=0.32\textwidth]{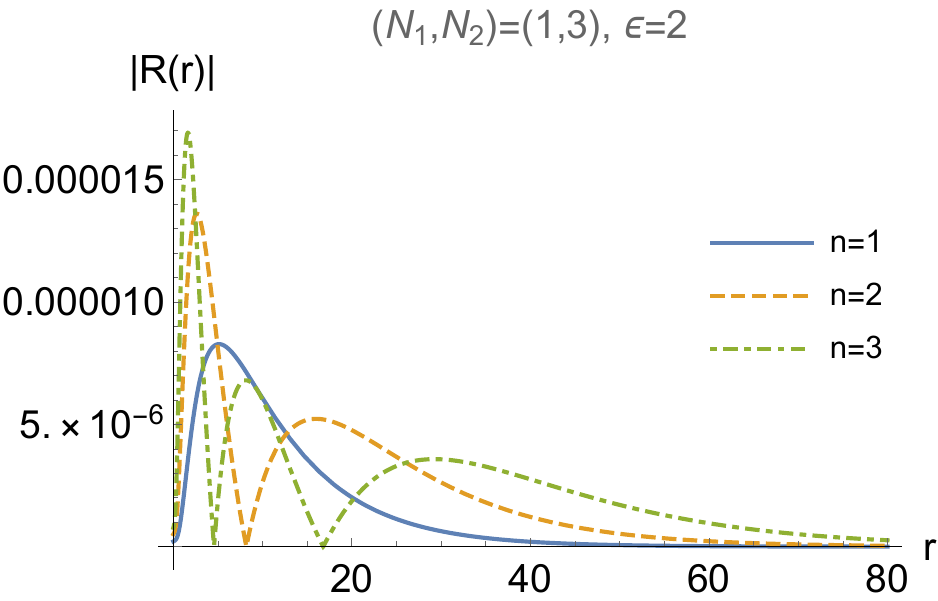}
  \includegraphics[width=0.32\textwidth]{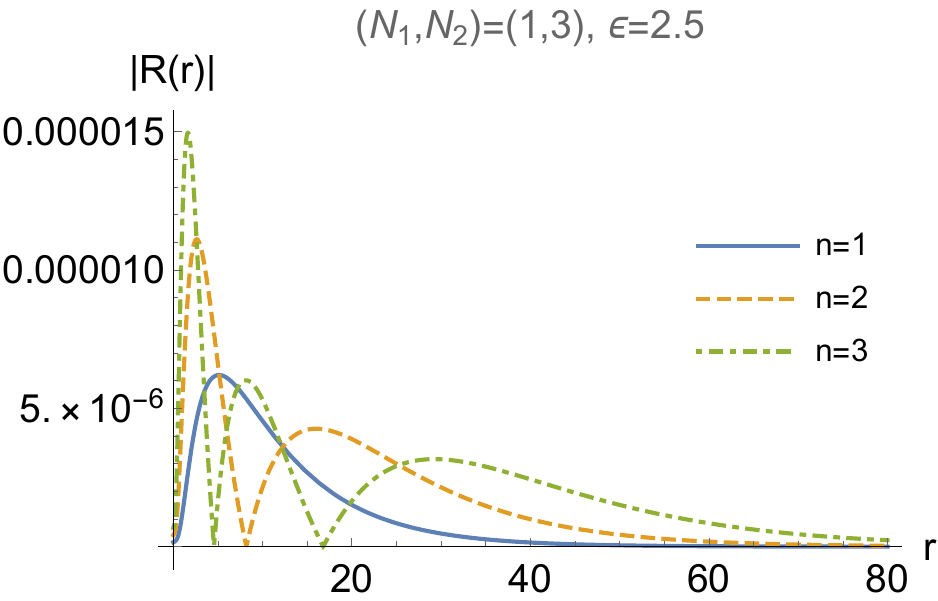}
  \includegraphics[width=0.32\textwidth]{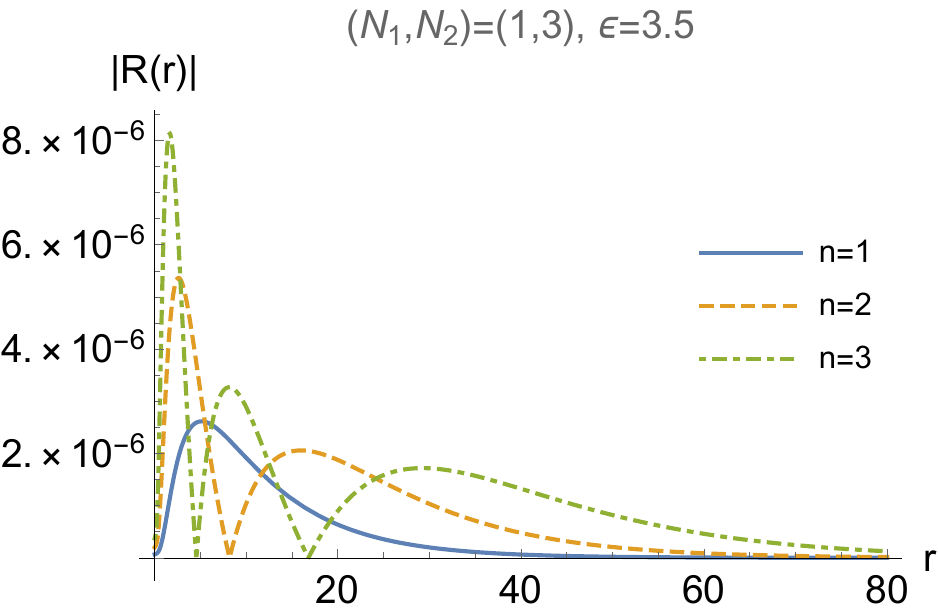}
\caption{Here we plot the normalized radial function $|R(r)|$ for low-lying quasi-bound states $(n=1,2,3)$ for black hole of various parameters. These states are characterised that the radial function falls off exponentially as $r\rightarrow \infty$. Fixed parameters are $m_p = \frac{2}{10}$, $q_1 = \frac{2}{10}$, $q_2=0$ and $Q_1=1$.}\label{epN}
\end{figure}
\begin{figure}[htpb]
  \centering
  \includegraphics[width=0.32\textwidth]{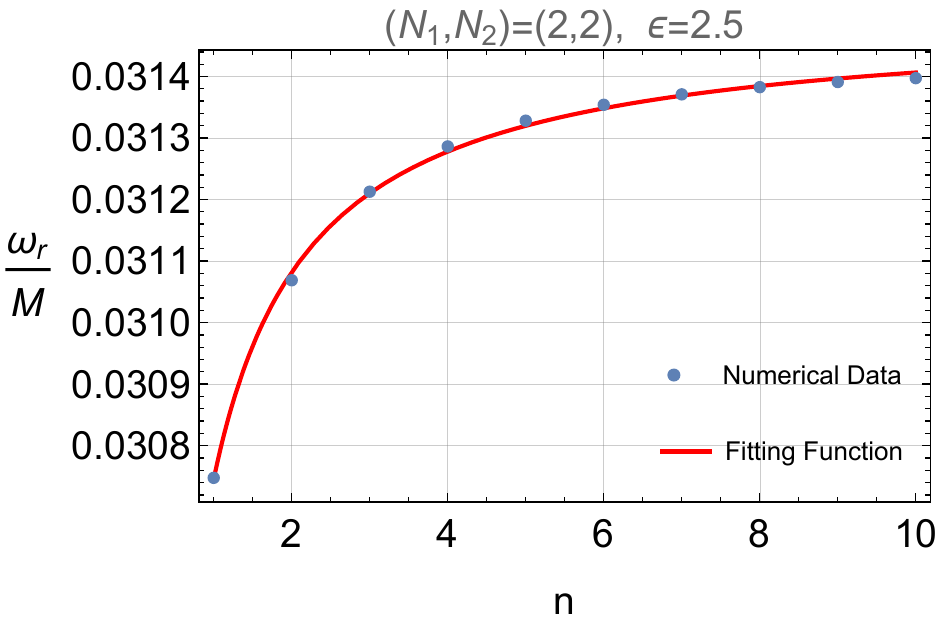}
  \includegraphics[width=0.32\textwidth]{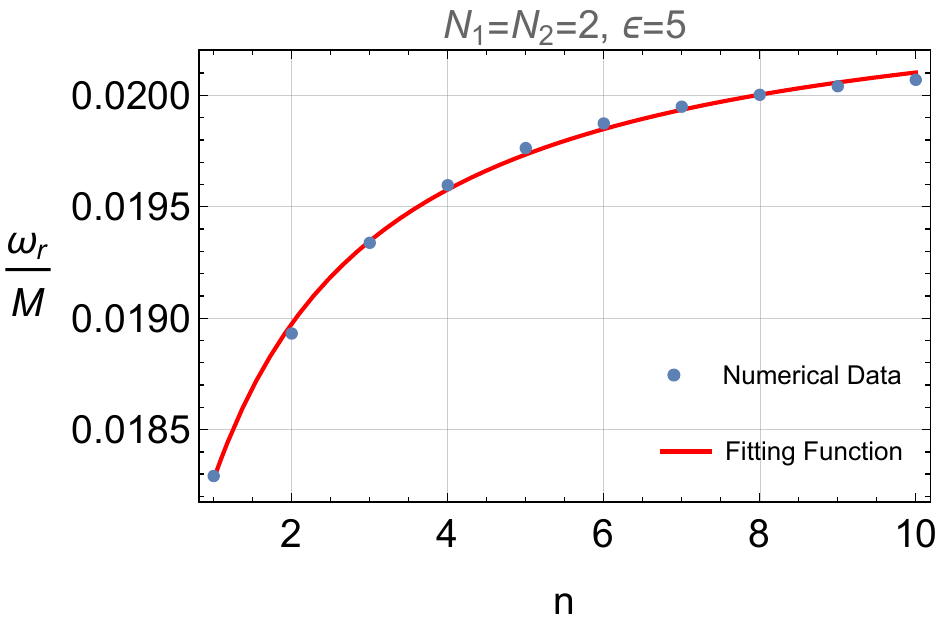}
  \includegraphics[width=0.32\textwidth]{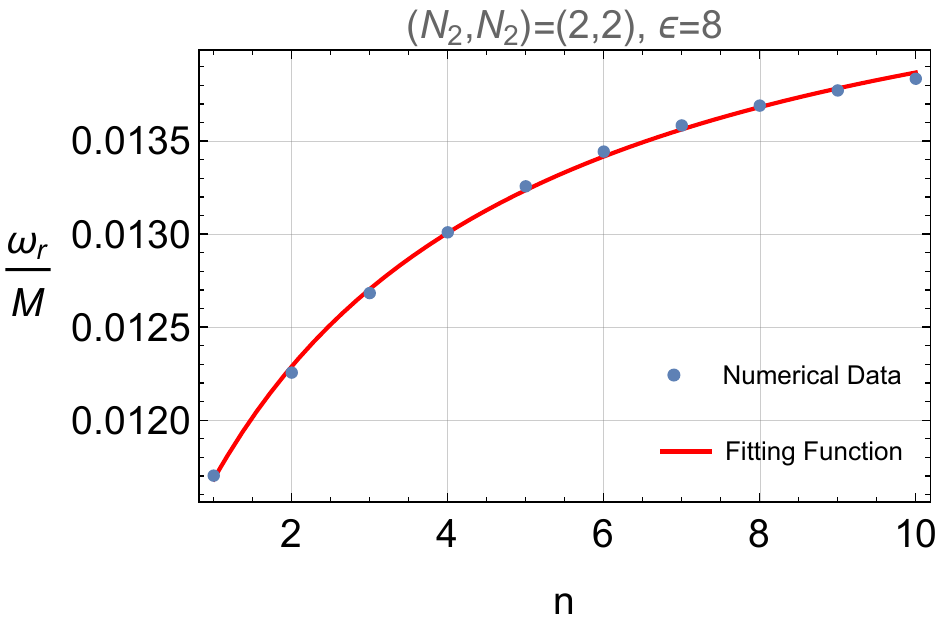}
  \includegraphics[width=0.32\textwidth]{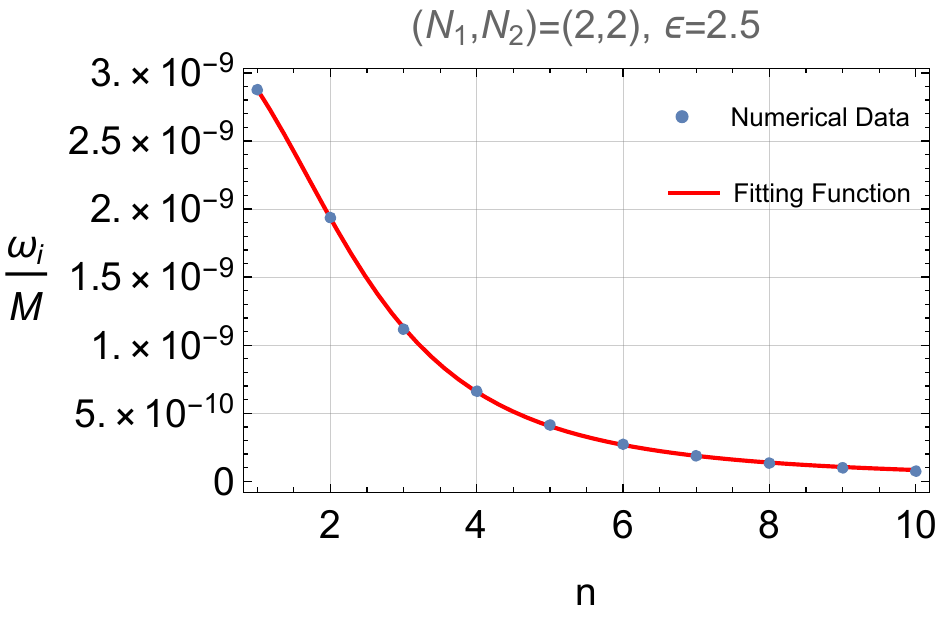}
  \includegraphics[width=0.32\textwidth]{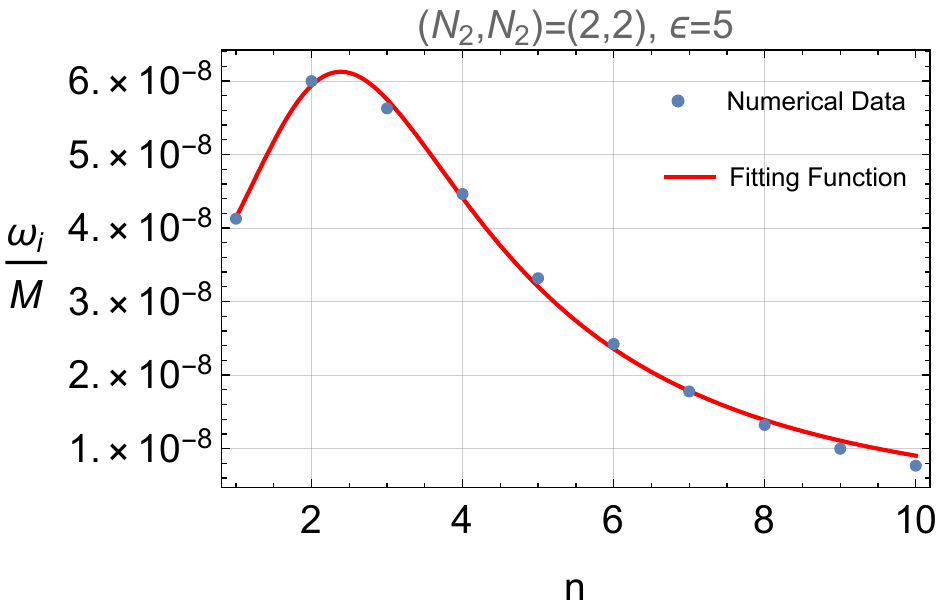}
  \includegraphics[width=0.32\textwidth]{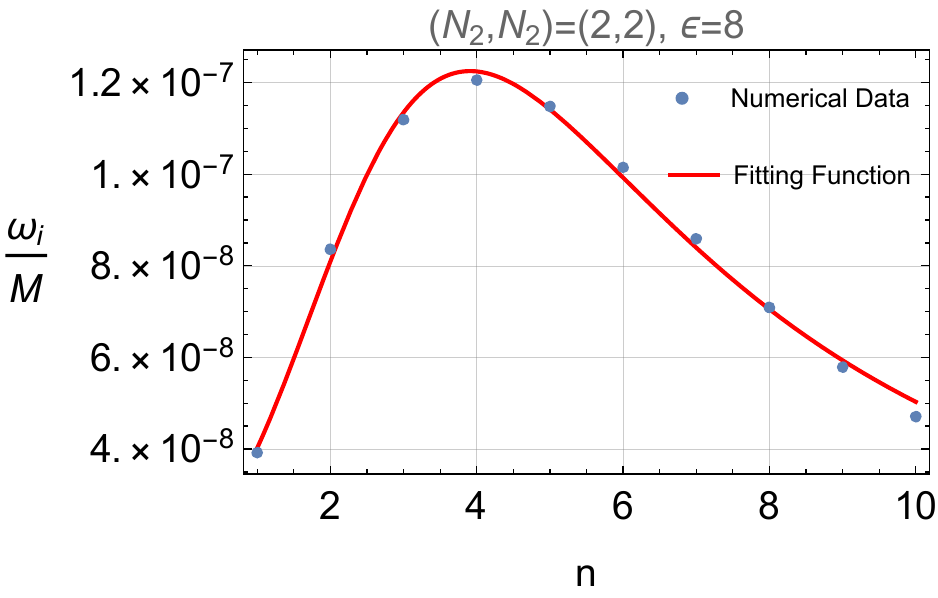}
\caption{Here we plot the $\omega_r/M$ and $\omega_i/M$ with respect to the overtone $n$ for various $\epsilon$, for the case of $(N_1, N_2)=(2,2)$. The solid lines are produced by the corresponding fitting functions in \eqref{fitf1}. We see that our fitting functions match the data remarkably well. Fixed parameters are $m_p = \frac{2}{10}$, $q_1 = \frac{2}{10}$, $q_2=0$ and $Q_1=1$.}\label{22vN1}
\end{figure}
\begin{figure}[htpb]
  \includegraphics[width=0.32\textwidth]{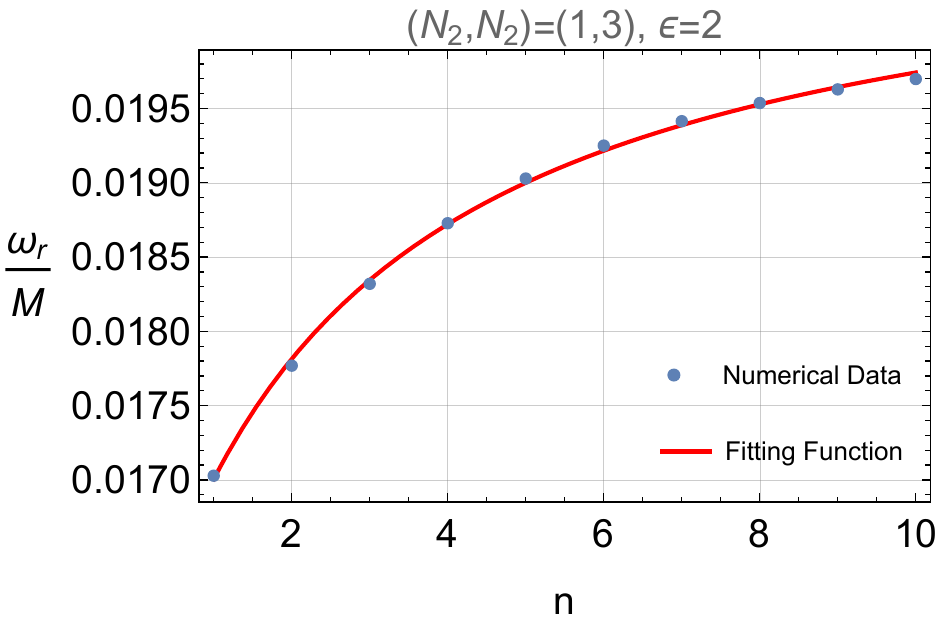}
  \includegraphics[width=0.32\textwidth]{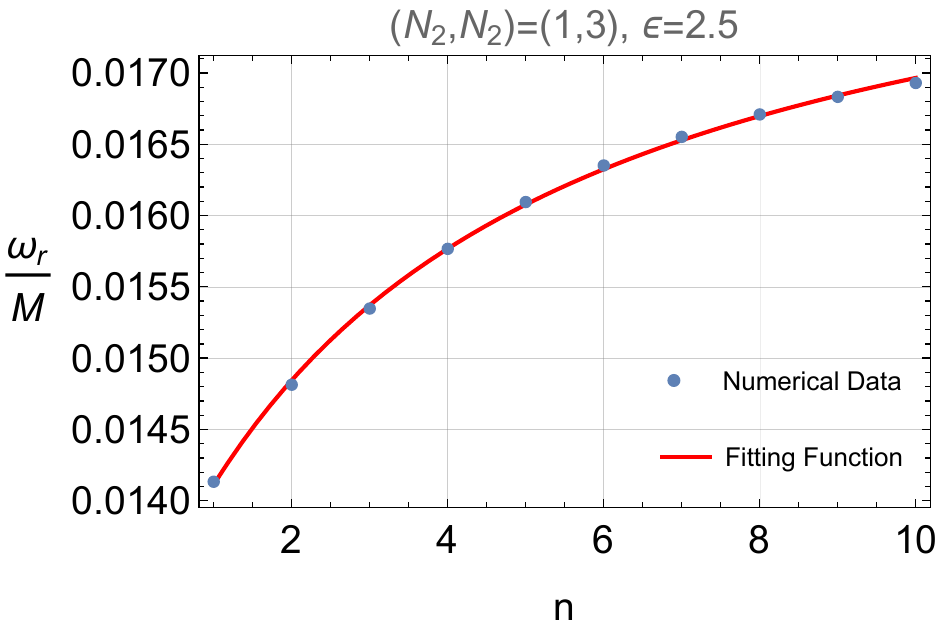}
  \includegraphics[width=0.32\textwidth]{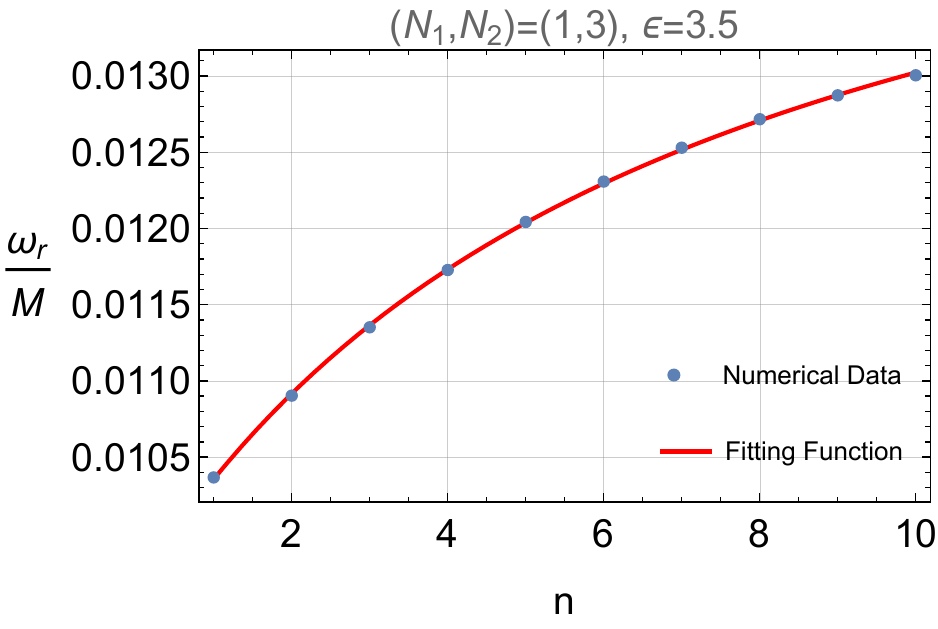}\\ \includegraphics[width=0.32\textwidth]{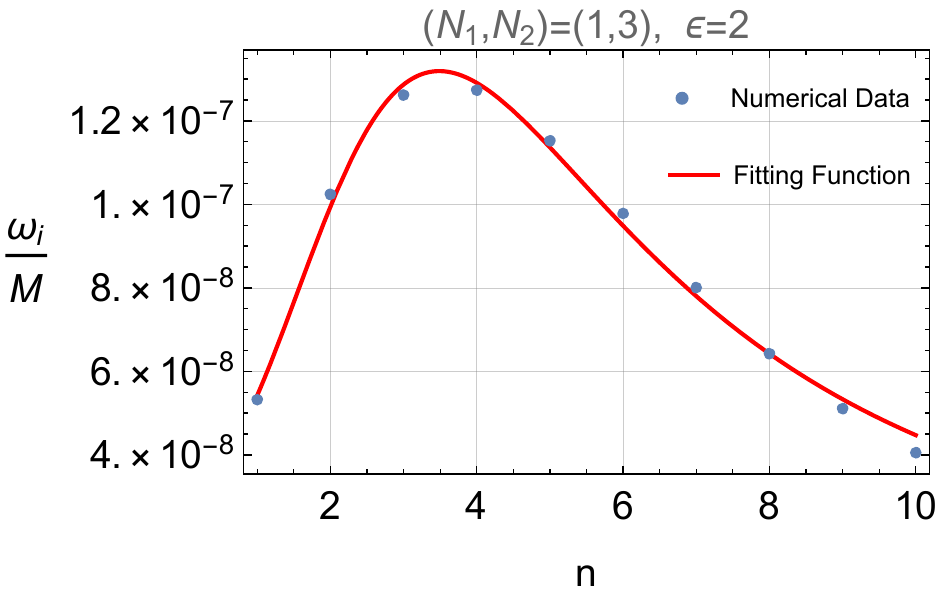}
  \includegraphics[width=0.32\textwidth]{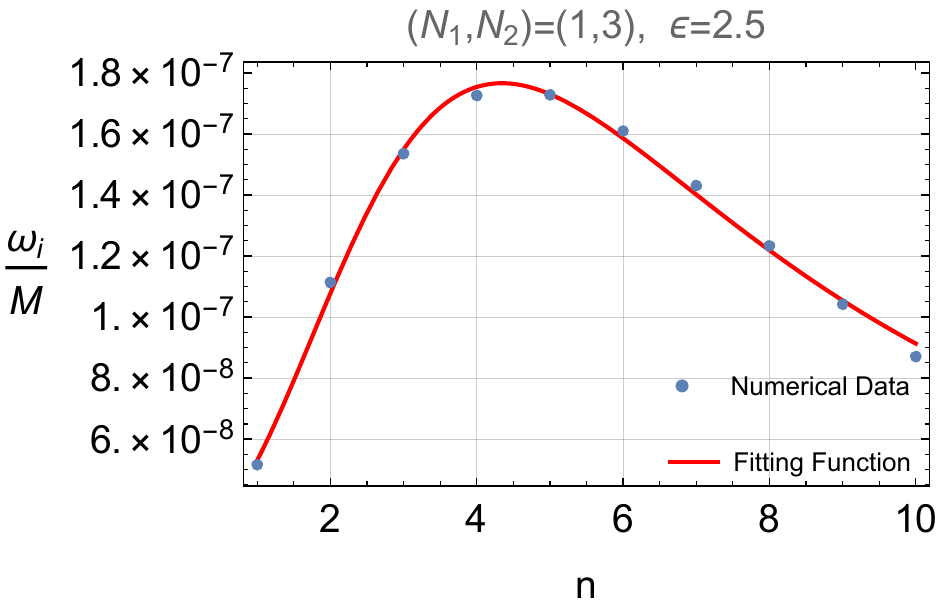}
  \includegraphics[width=0.32\textwidth]{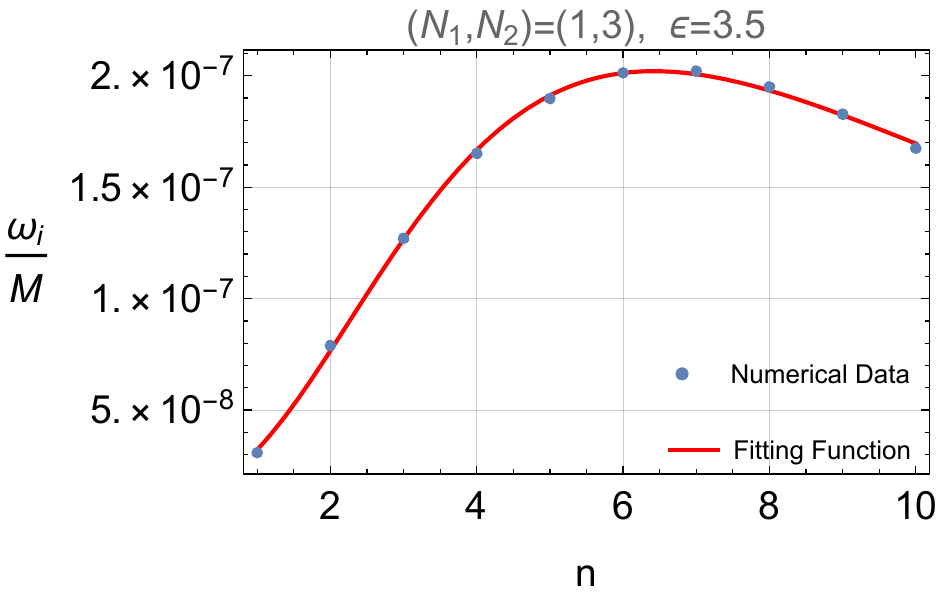}
 \caption{Here we plot the $\omega_r/M$ and $\omega_i/M$ with respect to the overtone $n$ for various $\epsilon$, for the case of $(N_1, N_2)=(1,3)$. The solid lines are produced by the corresponding fitting functions in \eqref{fitf1}. We see that our fitting functions match the data remarkably well. Fixed parameters are $m_p = \frac{2}{10}$, $q_1 = \frac{2}{10}$, $q_2=0$ and $Q_1=1$.}\label{22vN2}
\end{figure}

\begin{figure}[htpb]
  \centering
  \includegraphics[width=0.4\textwidth]{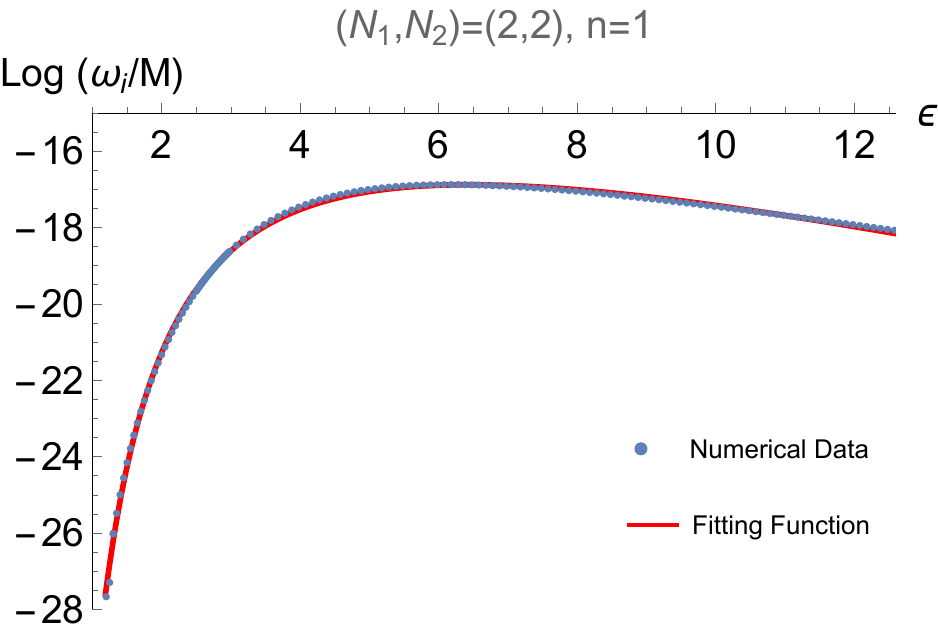}
  \includegraphics[width=0.4\textwidth]{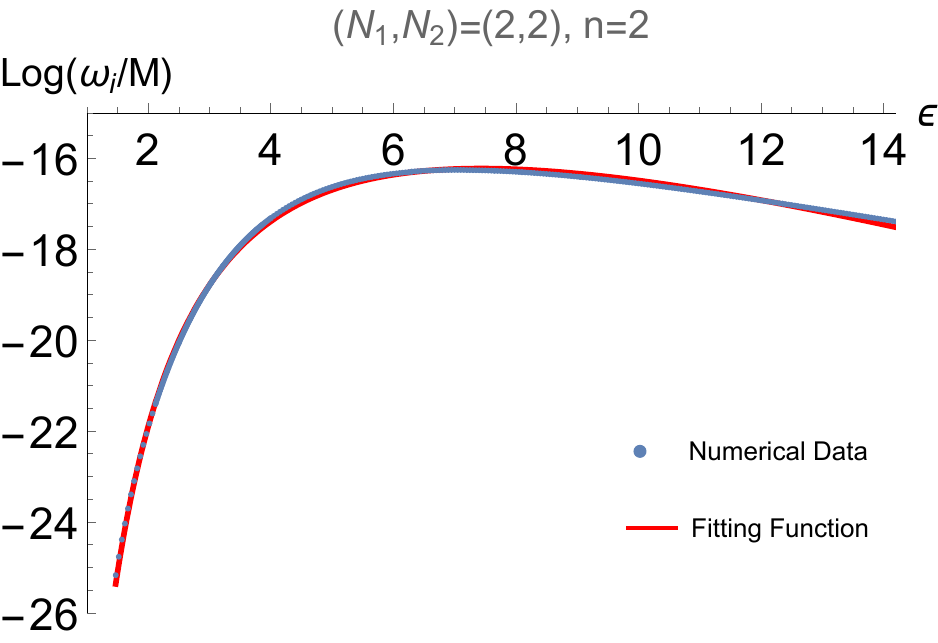}
  \includegraphics[width=0.4\textwidth]{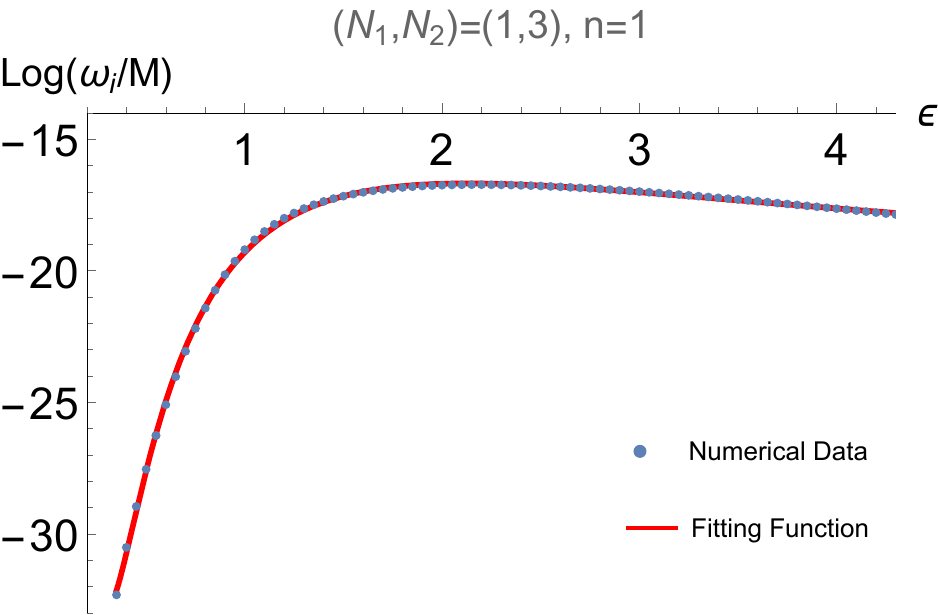}
  \includegraphics[width=0.4\textwidth]{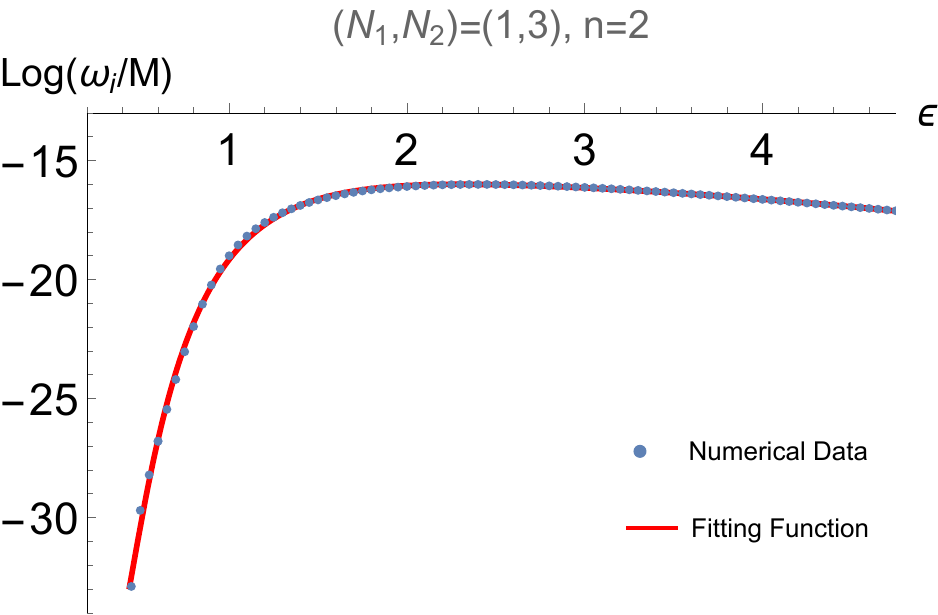}
 \caption{The plot of  $\log (\omega_i/M)$ with respect to $\epsilon$ in $(N_1, N_2)=(2, 2)$ and $(N_1, N_2)=(1, 3)$ examples.  Fixed parameters are $m_p = \frac{2}{10}$, $q_1 = \frac{2}{10}$, $q_2=0$ and $Q_1=1$. }\label{epome}
\end{figure}

\begin{figure}[hbtp]
  \centering
  \includegraphics[width=0.4\textwidth]{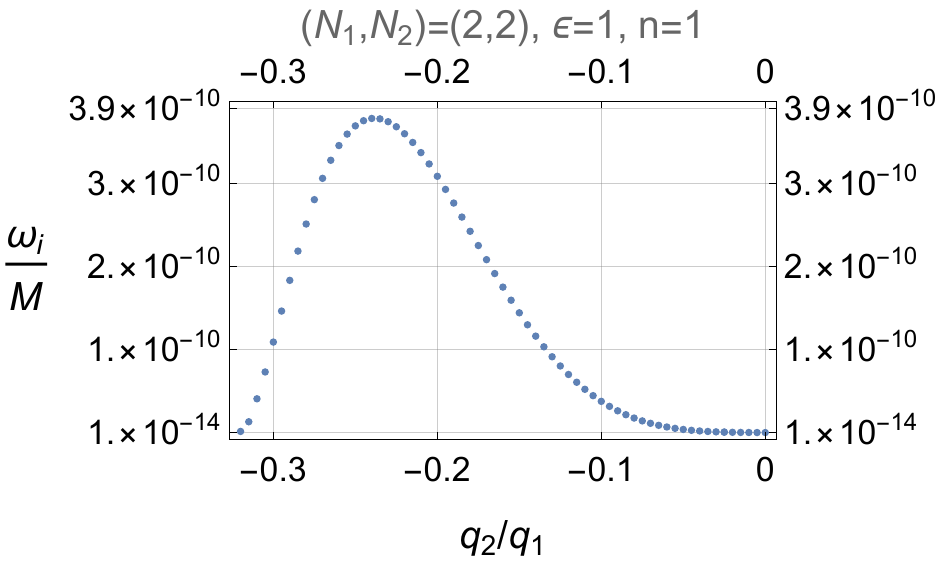}
  \includegraphics[width=0.4\textwidth]{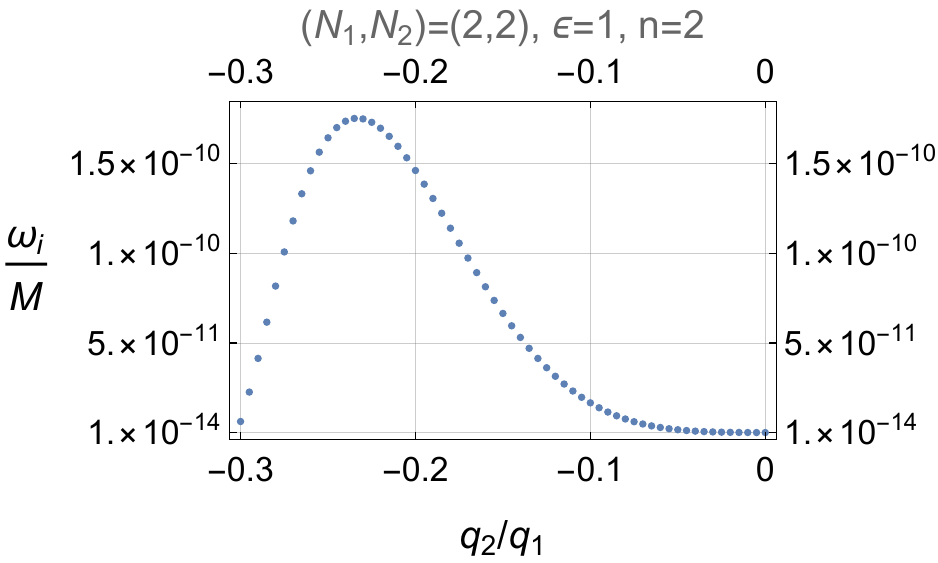}\\
  $\phantom{xxx}$\\
  \includegraphics[width=0.4\textwidth]{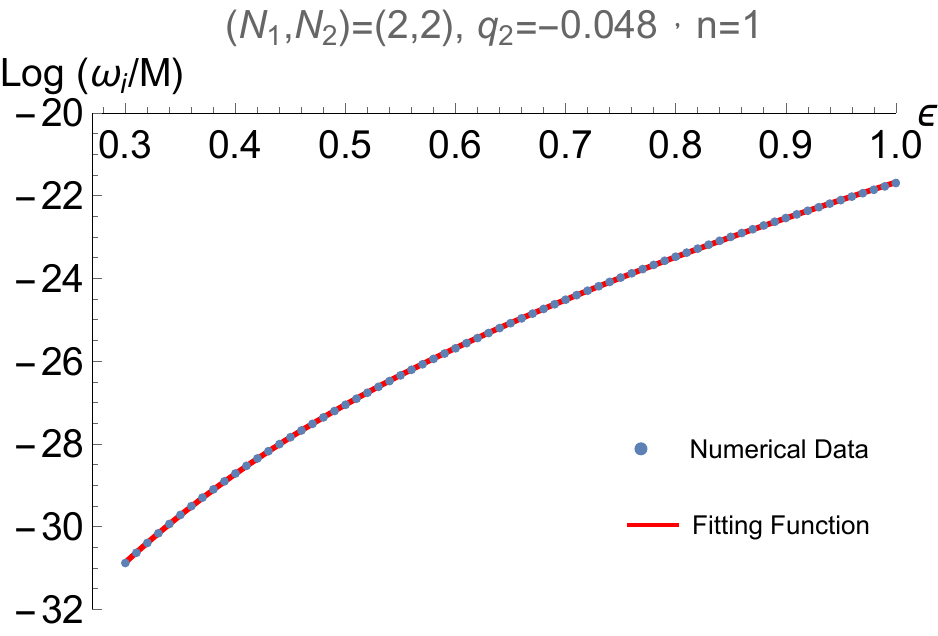}
  \includegraphics[width=0.4\textwidth]{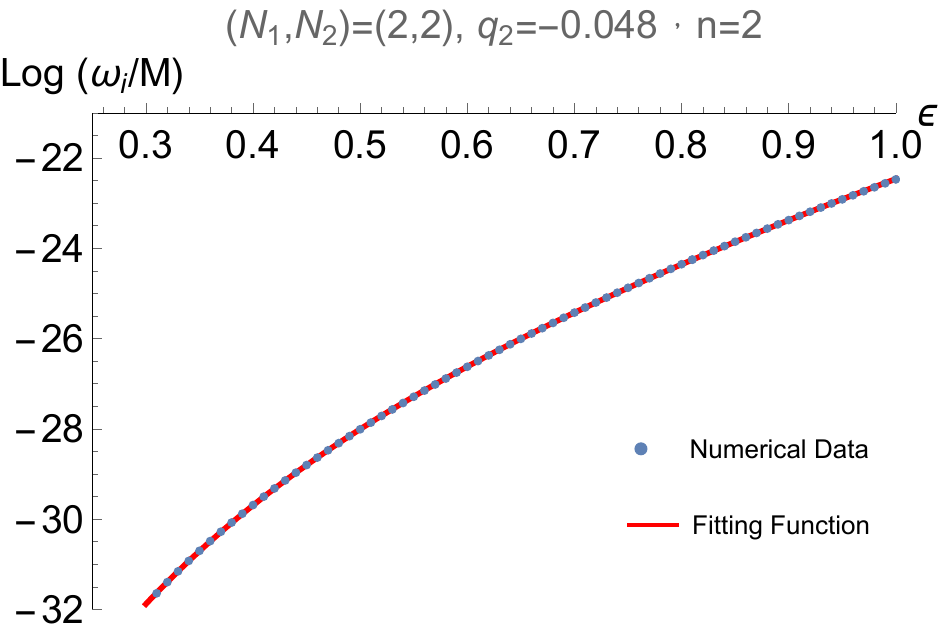}
 \caption{In the first row, we plot the $\omega_i/M$ in terms of the fundamental charge ratio $q_2/q_1$, with fixed parameters  $m_p = \frac{2}{10}$, $q_1 = \frac{2}{10}$, $Q_1=1$ and $\epsilon=1$. We see that $\epsilon=1$ state can actually become quite unstable as we run $q_2$ towards to some negative value, with $q_2=-0.048$ gives the maximum instability of order $10^{-10}$. In the second row, we therefore choosing $q_2 = -0.048$, we plot the $\log \omega_i/M$ with respect to $\epsilon$. This establishes that the $\epsilon=0.3$ black hole is also unstable.}\label{q2ome}
\end{figure}
\begin{figure}[htpb]
  \centering
  \includegraphics[width=0.4\textwidth]{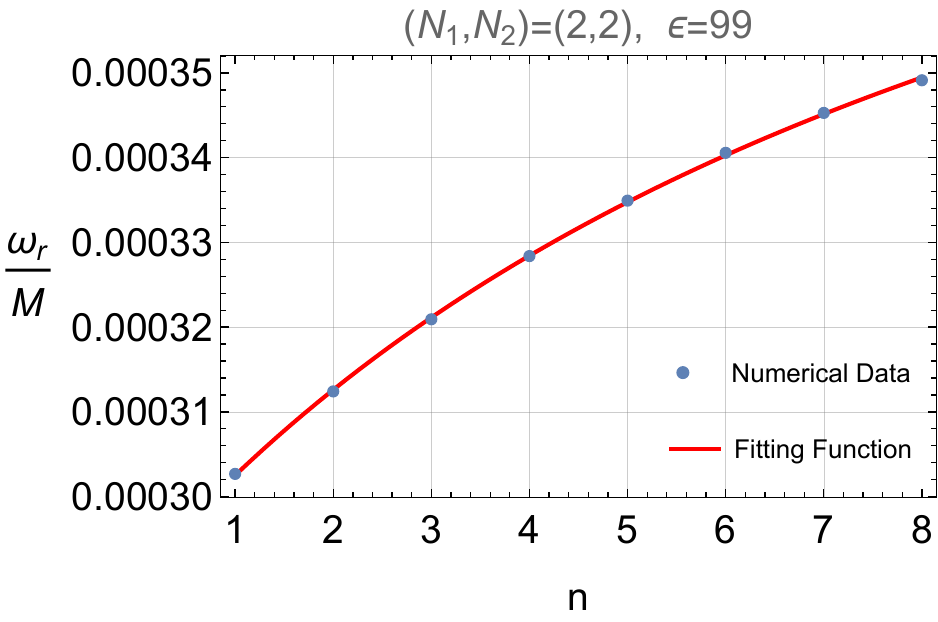}
  \includegraphics[width=0.4\textwidth]{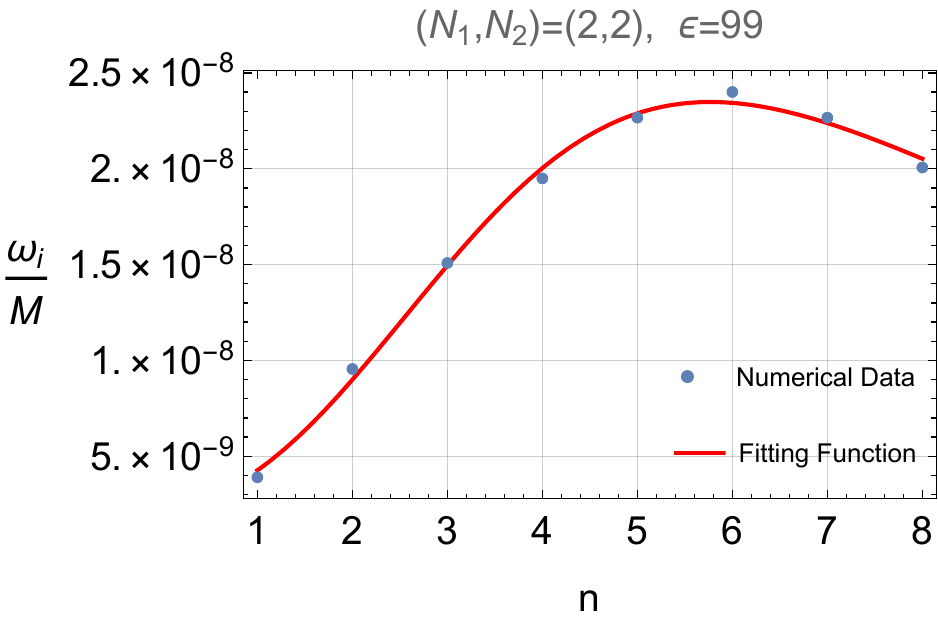}
  \includegraphics[width=0.4\textwidth]{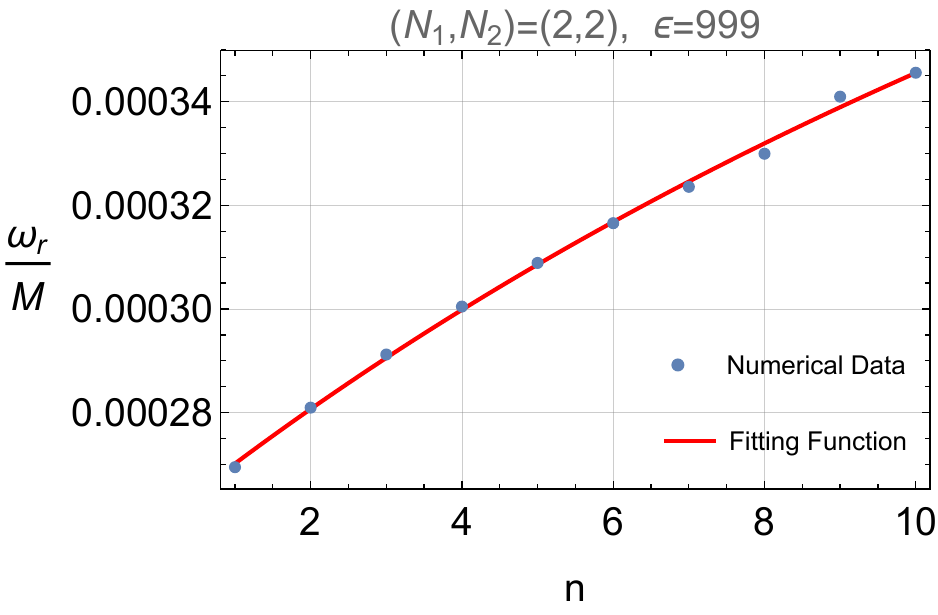}
  \includegraphics[width=0.4\textwidth]{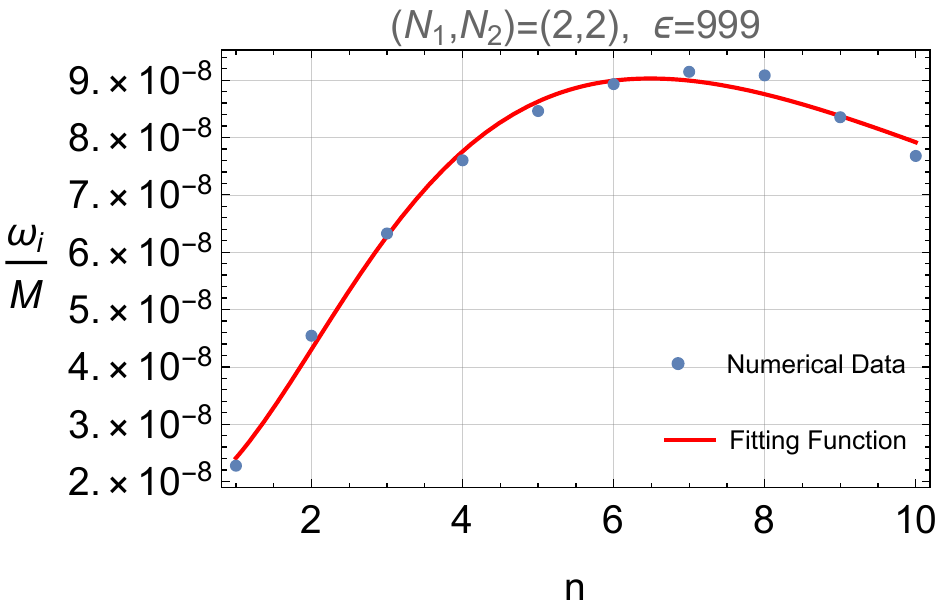}
\caption{For $(N_1, N_2)=(2,2)$, we give the $\omega_r/M$ and $\omega_i/M$ with respect to $n$ for $Q_2 = 1$ $(\epsilon=99)$ and $0.1$ $(\epsilon=999)$, with fixed $m_p = \frac{5}{100}$, $ q_1 = q_2 = \frac{2}{100}$, $Q_2 = 100$. The solid lines are produced by the fitting functions in \eqref{fitf1}.}\label{22vDN1}
\end{figure}
\begin{figure}[htpb]
  \centering
  \includegraphics[width=0.4\textwidth]{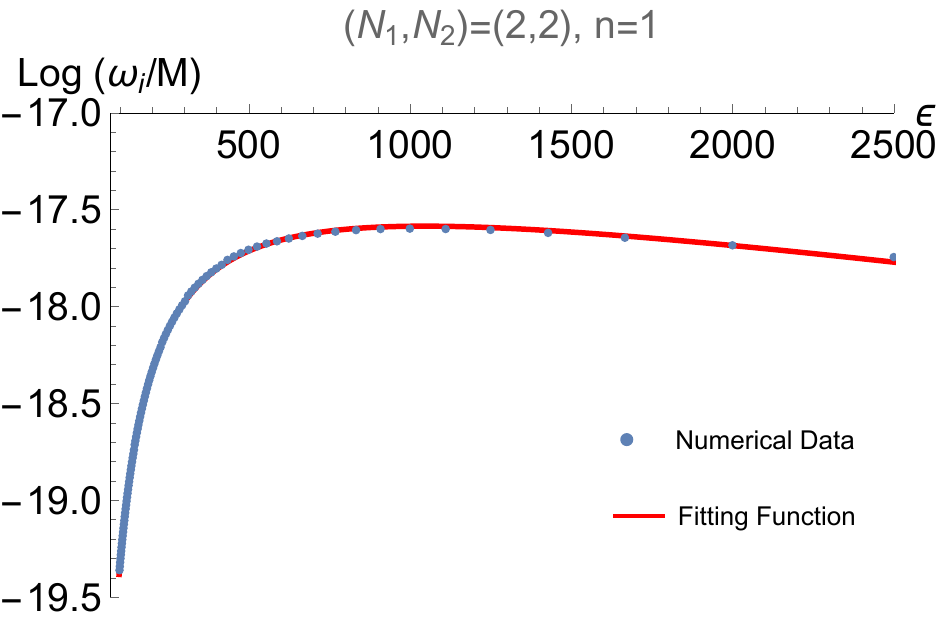}
  \includegraphics[width=0.4\textwidth]{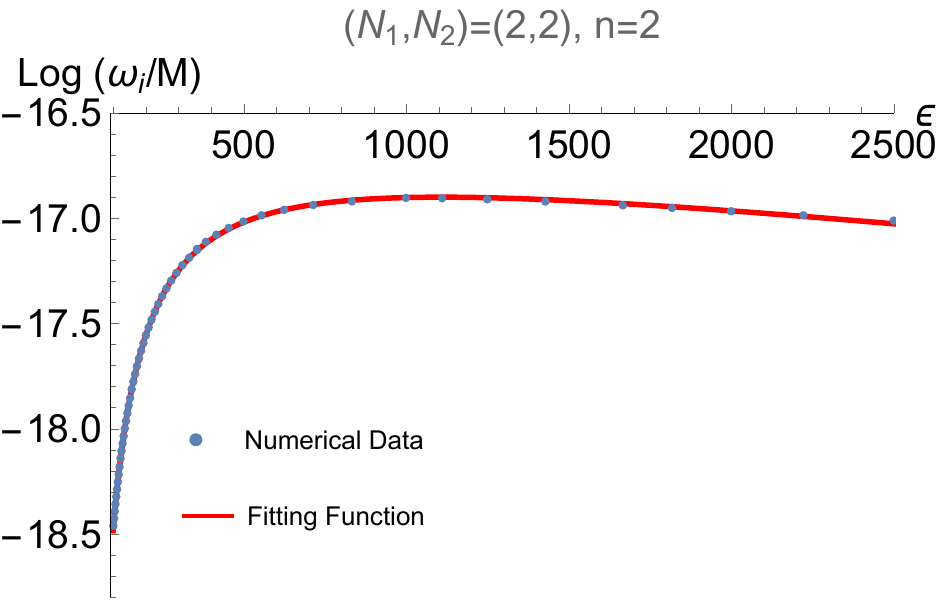}
 \caption{For $(N_1, N_2)=(2,2)$, we plot $\log (\omega_i/M)$ as a function of $\epsilon$, with fixed $m_p = \frac{5}{100}$, $ q_1 = q_2 = \frac{2}{100}$, $Q_2 = 100$. The dimensionless parameter $\epsilon$ runs from $99$ to $2499$, corresponding to $Q_1$ running from 1 to 0.04. The solid lines drawn from \eqref{fitf22} match the numerical data remarkably well for such large expanse of $\epsilon$ value.}\label{epomeD}
\end{figure}

\newpage

\providecommand{\href}[2]{#2}\begingroup\raggedright


\begin{thebibliography}{99}

\bibitem{LIGOScientific:2016aoc}
{\scshape LIGO Scientific, Virgo} collaboration, B.P. Abbott et~al.,
  \emph{{Observation of gravitational waves from a binary black hole merger}},
  \href{http://dx.doi.org/10.1103/PhysRevLett.116.061102}{\emph{Phys. Rev.
  Lett.} {\bf 116} (2016) 061102}, [\href{http://arxiv.org/abs/1602.03837}{{\tt
  1602.03837}}].

\bibitem{EventHorizonTelescope:2019dse}
{\scshape Event Horizon Telescope} collaboration, K.~Akiyama et~al.,
  \emph{{First M87 event horizon telescope results. I. The shadow of the
  supermassive black hole}},
  \href{http://dx.doi.org/10.3847/2041-8213/ab0ec7}{\emph{Astrophys. J. Lett.}
  {\bf 875} (2019) L1}, [\href{http://arxiv.org/abs/1906.11238}{{\tt
  1906.11238}}].

\bibitem{Press:1972zz}
W.H. Press and S.~A. Teukolsky, \emph{{Floating orbits, superradiant
  scattering and the black-hole Bomb}},
  \href{http://dx.doi.org/10.1038/238211a0}{\emph{Nature} {\bf 238} (1972)
  211--212}.

\bibitem{Bekenstein:1973mi}
J.D. Bekenstein, \emph{{Extraction of energy and charge from a black hole}},
  \href{http://dx.doi.org/10.1103/PhysRevD.7.949}{\emph{Phys. Rev. D} {\bf 7}
  (1973) 949--953}.

\bibitem{Teukolsky:1974yv}
S.A. Teukolsky and W.H. Press, \emph{{Perturbations of a rotating black hole.
  III - Interaction of the hole with gravitational and electromagnet ic
  radiation}}, \href{http://dx.doi.org/10.1086/153180}{\emph{Astrophys. J.}
  {\bf 193} (1974) 443--461}.

\bibitem{Starobinskil:1974nkd}
A.A. Starobinskil and S.M. Churilov, \emph{{Amplification of electromagnetic
  and gravitational waves scattered by a rotating ''black hole''}}, {\emph{Sov.
  Phys. JETP} {\bf 65} (1974) 1--5}.

\bibitem{Starobinsky:1973aij}
A.A. Starobinsky, \emph{{Amplification of waves reflected from a rotating
  ``black hole''.}}, {\emph{Sov. Phys. JETP} {\bf 37} (1973) 28--32}.

\bibitem{Brito:2015oca}
R.~Brito, V.~Cardoso and P.~Pani, \emph{{Superradiance}: {New frontiers in
  black hole physics}},
  \href{http://dx.doi.org/10.1007/978-3-319-19000-6}{\emph{Lect. Notes Phys.}
  {\bf 906} (2015) pp.1--237}, [\href{http://arxiv.org/abs/1501.06570}{{\tt
  1501.06570}}].

\bibitem{Damour:1976kh}
T.~Damour, N.~Deruelle and R.~Ruffini, \emph{{On quantum resonances in
  stationary geometries}},
  \href{http://dx.doi.org/10.1007/BF02725534}{\emph{Lett. Nuovo Cim.} {\bf 15}
  (1976) 257--262}.

\bibitem{Detweiler:1980uk}
S.~L. Detweiler, \emph{{Klein-Gordon equation and rotating black holes}},
  \href{http://dx.doi.org/10.1103/PhysRevD.22.2323}{\emph{Phys. Rev. D} {\bf
  22} (1980) 2323--2326}.

\bibitem{Furuhashi:2004jk}
H.~Furuhashi and Y.~Nambu, \emph{{Instability of massive scalar fields in
  Kerr-Newman space-time}},
  \href{http://dx.doi.org/10.1143/PTP.112.983}{\emph{Prog. Theor. Phys.} {\bf
  112} (2004) 983--995}, [\href{http://arxiv.org/abs/gr-qc/0402037}{{\tt
  gr-qc/0402037}}].

\bibitem{Dolan:2007mj}
S.R. Dolan, \emph{{Instability of the massive Klein-Gordon field on the Kerr
  spacetime}}, \href{http://dx.doi.org/10.1103/PhysRevD.76.084001}{\emph{Phys.
  Rev. D} {\bf 76} (2007) 084001}, [\href{http://arxiv.org/abs/0705.2880}{{\tt
  0705.2880}}].

\bibitem{Huang:2018qdl}
Y.~Huang, D.-J. Liu, X.-h. Zhai and X.-z. Li, \emph{{Instability for massive
  scalar fields in Kerr-Newman spacetime}},
  \href{http://dx.doi.org/10.1103/PhysRevD.98.025021}{\emph{Phys. Rev. D} {\bf
  98} (2018) 025021}, [\href{http://arxiv.org/abs/1807.06263}{{\tt
  1807.06263}}].

\bibitem{Cardoso:2004hs}
V.~Cardoso and O.J.C. Dias, \emph{{Small Kerr-anti-de Sitter black holes are
  unstable}}, \href{http://dx.doi.org/10.1103/PhysRevD.70.084011}{\emph{Phys.
  Rev. D} {\bf 70} (2004) 084011},
  [\href{http://arxiv.org/abs/hep-th/0405006}{{\tt hep-th/0405006}}].

\bibitem{Cardoso:2004nk}
V.~Cardoso, O.J.C. Dias, J.P.S. Lemos and S.~Yoshida, \emph{{The black hole
  bomb and superradiant instabilities}},
  \href{http://dx.doi.org/10.1103/PhysRevD.70.049903}{\emph{Phys. Rev. D} {\bf
  70} (2004) 044039}, [\href{http://arxiv.org/abs/hep-th/0404096}{{\tt
  hep-th/0404096}}].

\bibitem{Li:2012rx}
R.~Li, \emph{{Superradiant instability of charged massive scalar field in
  Kerr-Newman-anti-de Sitter black hole}},
  \href{http://dx.doi.org/10.1016/j.physletb.2012.07.015}{\emph{Phys. Lett. B}
  {\bf 714} (2012) 337--341}, [\href{http://arxiv.org/abs/1205.3929}{{\tt
  1205.3929}}].

\bibitem{Wang:2014eha}
M.~Wang and C.~Herdeiro, \emph{{Superradiant instabilities in a D-dimensional
  small Reissner-Nordstr\"om-anti-de Sitter black hole}},
  \href{http://dx.doi.org/10.1103/PhysRevD.89.084062}{\emph{Phys. Rev. D} {\bf
  89} (2014) 084062}, [\href{http://arxiv.org/abs/1403.5160}{{\tt 1403.5160}}].

\bibitem{Bosch:2016vcp}
P.~Bosch, S.R. Green and L.~Lehner, \emph{{Nonlinear evolution and final fate
  of charged anti\textendash{}de Sitter black hole superradiant instability}},
  \href{http://dx.doi.org/10.1103/PhysRevLett.116.141102}{\emph{Phys. Rev.
  Lett.} {\bf 116} (2016) 141102}, [\href{http://arxiv.org/abs/1601.01384}{{\tt
  1601.01384}}].

\bibitem{Ganchev:2016zag}
B.~Ganchev, \emph{{Superradiant instability in AdS}},
  \href{http://arxiv.org/abs/1608.01798}{{\tt 1608.01798}}.

\bibitem{Hod:2012wmy}
S.~Hod, \emph{{Stability of the extremal Reissner-Nordstroem black hole to
  charged scalar perturbations}},
  \href{http://dx.doi.org/10.1016/j.physletb.2012.06.043}{\emph{Phys. Lett. B}
  {\bf 713} (2012) 505--508}, [\href{http://arxiv.org/abs/1304.6474}{{\tt
  1304.6474}}].

\bibitem{Huang:2015jza}
J.-H. Huang and Z.-F. Mai, \emph{{Superradiantly stable non-extremal
  Reissner\textendash{}Nordstr\"om black holes}},
  \href{http://dx.doi.org/10.1140/epjc/s10052-016-4157-y}{\emph{Eur. Phys. J.
  C} {\bf 76} (2016) 314}, [\href{http://arxiv.org/abs/1503.01221}{{\tt
  1503.01221}}].

\bibitem{Hod:2015hza}
S.~Hod, \emph{{Stability of highly-charged Reissner-Nordstr\"om black holes to
  charged scalar perturbations}},
  \href{http://dx.doi.org/10.1103/PhysRevD.91.044047}{\emph{Phys. Rev. D} {\bf
  91} (2015) 044047}, [\href{http://arxiv.org/abs/1504.00009}{{\tt
  1504.00009}}].

\bibitem{Duff:1995sm}
M.J.~Duff, J.T.~Liu and J.~Rahmfeld,
\emph{Four-dimensional string-string-string triality,}
Nucl. Phys. B \textbf{459}, 125-159 (1996)
doi:10.1016/0550-3213(95)00555-2
[arXiv:hep-th/9508094 [hep-th]].

\bibitem{Cvetic:1999xp}
M.~Cveti\v c, M.J.~Duff, P.~Hoxha, J.T.~Liu, H.~L\"u, J.X.~Lu, R.~Martinez-Acosta, C.N.~Pope, H.~Sati and T.~A.~Tran,
\emph{Embedding AdS black holes in ten-dimensions and eleven-dimensions,}
Nucl. Phys. B \textbf{558}, 96-126 (1999)
doi:10.1016/S0550-3213(99) 00419-8
[arXiv:hep-th/9903214 [hep-th]].

\bibitem{Lu:2013eoa}
H.~L\"u, \emph{{Charged dilatonic ads black holes and magnetic AdS$_{D-2} \times
  R^{2}$ vacua}}, \href{http://dx.doi.org/10.1007/JHEP09(2013)112}{\emph{JHEP}
  {\bf 09} (2013) 112}, [\href{http://arxiv.org/abs/1306.2386}{{\tt
  1306.2386}}].

\bibitem{DiMenza:2014vpa}
L.~Di~Menza and J.-P. Nicolas, \emph{{Superradiance on the
  Reissner\textendash{}Nordstr\o{}m metric}},
  \href{http://dx.doi.org/10.1088/0264-9381/32/14/145013}{\emph{Class. Quant.
  Grav.} {\bf 32} (2015) 145013}, [\href{http://arxiv.org/abs/1411.3988}{{\tt
  1411.3988}}].

\end{thebibliography}
\end{document}